\documentstyle[sprocl]{article}

\def\be{\begin{equation}}
\def\ee{\end{equation}}
\def\bea{\begin{eqnarray}}
\def\eea{\end{eqnarray}}

\begin{document}

\hfill UMD-PP-98-71

\bigskip

\bigskip

\title{ SUPERSYMMETRIC GRAND UNIFICATION:\\ Lectures at TASI97 }

\author{ R. N. MOHAPATRA}

\address{Department of Physics, University of Maryland, College Park
\\ MD 20742, USA\\E-mail: rmohapatra@umdhep.umd.edu}

\maketitle\abstracts{ 
One of the key ingredients of physics beyond the standard model is widely
believed to be a symmetry between the fermions and the bosons known as
supersymmetry. The reason for this is that the milder divergence structure 
of field theories with this symmetry may explain why the electroweak scale (or
the Higgs mass ) is stable under radiative corrections. Two other reasons
adding to this belief are : (i) a way to understand the origin
of the electroweak symmetry breaking as a consequence of radiative corrections
and (ii) the particle content of the minimal supersymmetric model that leads in
a natural way to the unification of the three gauge couplings of the standard 
model at a high scale. This last observation suggests that at scales 
close to
the Planck scale, all matter and all forces may unify into a single matter 
and a single force leading to a supersymmetric grand unified theory. It is the
purpose of these lectures to provide a pedagogical discussion of the various
kinds of supersymmetric unified theories beyond the minimal supersymmetric
standard model (MSSM) including SUSY GUTs and present a brief overview of 
their implications. Questions 
such as proton decay, R-parity violation, doublet triplet splitting etc.
are discussed. Exhaustive discussion of $SU(5)$ and $SO(10)$ models and 
less detailed ones
for other GUT models such as those based on $E_6$, $SU(5)\times SU(5)$,
flipped SU(5) and $SU(6)$ are presented. }

\newpage

\noindent {\Large Table of contents}

\bigskip

\noindent {\bf Lecture I: Introduction}

\begin{quote}
1.1  Brief introduction to supersymmetric field theories\\

1.2  The minimal supersymmetric standard model (MSSM)\\

1.3  Why go beyond the MSSM ?\\

1.4  Origin of supersymmetry breaking: gravity mediated, gauge mediated
and U(1)-mediated models\\

1.5  Supersymmetric left-right models

\end{quote}

\noindent {\bf Lecture II : Unification of couplings}

\begin{quote}

2.1  Unification of gauge couplings (UGC)\\

2.2  UGC with intermediate scales\\

2.3  Yukawa unification and $m_b, m_{\tau}$ and $m_{t}$

\end{quote}

\noindent {\bf Lecture III. Supersymmetric SU(5)}

\begin{quote}

3.1  Fermion and Higgs assignment, symmetry breaking\\

3.2  Low energy spectrum and doublet-triplet splitting\\

3.3  Fermion masses and  Proton decay\\

3.4  Other implications of SU(5)\\

3.5  Problems and prospects

\end{quote}

\noindent {\bf Lecture IV: Left-right symmetric GUT: The SO(10) example}

\begin{quote}

4.1  Group theory of SO(10)\\

4.2  Symmetry breaking and fermion masses\\

4.3  Neutrino masses and R-parity breaking\\

4.4  Doublet-triplet splitting\\

4.5  Final comments 

\end{quote}

\noindent{\bf Lecture V: Beyond simple SUSY GUT's- four examples}

\begin{quote}

5.1  $E_6$\\

5.2  $SU(5)\times SU(5)$ \\

5.3  $SU(5)\times U(1)$\\

5.4  $SU(6)$ and new approach to doublet-triplet splitting

\end{quote}   

\noindent{\bf Epilogue}
                  
\section{Introduction}

One of the fundamental new symmetries of nature that has been the
subject of intense discussion in particle physics of the past decade
is the symmetry between bosons and fermions, known as
supersymmetry. This symmetry was introduced in the early 1970's by
Golfand, Likhtman, Akulov, Volkov, Wess and Zumino. In addition to
the obvious fact that it provides the hope of an unified
understanding of the two known forms of matter, the bosons and
fermions, it has also provided a mechanism to solve two conceptual
problems of the standard model, viz. the possible origin of the
weak scale as well as its stability under quantum corrections. The
recent developments in strings, which embody supersymmetry in an
essential way also promise the fulfilment of the eternal dream of
all physicists to find an ultimate theory of everything. It would
thus appear that a large body of contemporary particle physicists
have accepted that the theory of particles and forces must incorporate
supersymmetry. This has important implications for
the nature of new physics beyond the standard model which is a
major focus of most research in particle physics at the moment. Ultimate test
of these ideas will of course come from the experimental discovery of
the superpartners of the standard model particles with masses under
a TeV and the standard model Higgs boson with mass less than about
$150$ GeV.

Since supersymmetry transforms a boson to a fermion and vice versa,
an irreducible representation of supersymmetry will contain in it   
both fermions and bosons. Therefore in a supersymmetric theory, all 
known particles are accompanied by a superpartner which is a fermion
if the known particle is a boson and vice versa. For instance, the
electron ($e$) supermultiplet will contain its superpartner
$\widetilde{e}$, (called the selectron) which has spin zero. 
 We will adopt the notation that the
superpartner of a particle will be denoted by the same symbol as the
particle with a `tilde' as above.
Furthermore, while supersymmetry does not commute with the Lorentz
transformations, it commutes with all internal symmetries; as a   
result, all non-Lorentzian quantum numbers for both the fermion and
boson in the same supermultiplet are the same. As in the case of
all symmetries realized in the Wigner-Weyl mode, 
in the limit of exact supersymmetry, all particles in the same
supermultiplet will have the same mass. Since this is contrary to
what is observed in nature, supersymmetry has to be a broken symmetry.
An interesting feature of supersymmetric theories is that
the supersymmetry breaking terms are fixed by the 
requirement that the mild divergence structure of the theory remains
uneffected. One then has a complete guide book for writing the
local field theories with broken supersymmetry. We will not discuss
the detailed introductory aspects of supersymmetry that are needed to write
the Lagrangian for these models and instead refer to books and
review articles on the subject\cite{ss|bagger,ss|mohap,ss|HaKa84,ss|Nil84}. 
Let us 
however give the bare outlines of how one goes about writing the
action for such models.

\subsection{Brief introduction to the supersymmetric field theories}

In order to write down the action for a supersymmetric field theory,
let us start by considering generic chiral fields denoted by
$\Phi(x, \theta)$ with component fields given by $(\phi,
\psi)$ and gauge fields denoted by $V(x, \theta,
\bar{\theta})$ with component gauge and gaugino fields given by
$(A^{\mu},\lambda)$. The action in the superfield notation is
	\begin{eqnarray}
S=\int d^4x \int d^2\theta d^2{\bar{\theta}} 
\Phi^{\dagger}e^V \Phi  + \int d^4x 
\int d^2\theta \left( W(\Phi) + W_{\lambda}(V) W_{\lambda}(V) \right)+h.c.
	\end{eqnarray}
In the above equation, the first term gives the gauge invariant
kinetic energy term for the matter fields $\Phi$; $W(\Phi)$ is a
holomorphic function of $\Phi$ and is called the superpotential; it
leads to the Higgs potential of the usual gauge field
theories. Secondly, $W_{\lambda}(V)\equiv {\cal D}^2\bar{{\cal D}}V$ where
${\cal D} \equiv \partial_\theta - i\sigma. \partial_x$, 
and the term involving $W_{\lambda}(V)$ leads to
the gauge invariant kinetic 
energy term for the gauge fields as well as for the gaugino
fields. In terms of the component fields the lagrangian can be
written as
	\begin{eqnarray}
{\cal L} = {\cal L}_{g} + {\cal L}_{matter} + {\cal L}_Y - V(\phi)
	\end{eqnarray}
where
	\begin{eqnarray}
{\cal L}_g = - \frac{1}{4}F^{\mu\nu}F_{\mu\nu} + \frac{1}{2}
\bar{\lambda}\gamma^{\mu} iD_\mu \lambda \nonumber\\ 
{\cal L}_{matter} = |D_\mu\phi|^2 + \bar{\psi}\gamma^\mu iD_\mu
\psi \nonumber\\ 
{\cal L}_V = \sqrt {2} g \bar\lambda \psi \phi^{\dagger} +\psi_a
\psi_b W_{ab} \nonumber\\ 
V(\phi) = |W_a|^2 + \frac{1}{2} {\cal D}_\alpha {\cal D}_\alpha
\label{ss.3}
	\end{eqnarray}
where $D_\mu$ stands for the covariant derivative with respect to
the gauge group and ${\cal D}_\alpha$ stands for the so-called
${\cal D}$-term and 
is given by ${\cal D}_\alpha= g \phi^\dagger T_\alpha \phi$ ($g$ is
the 
gauge coupling constant and $T_\alpha$ are the generators of the
gauge group).  $W_a$ and $W_{ab}$ are the first and second
derivative of the superpotential $W$ with respect to the superfield
with respect to the field $\Phi_a$, where the index $a$ stands for
different matter fields in the model.

A  very important property of supersymmetric field theories
is their ultraviolet behavior which have the extremely important
consequence that in the exact supersymmetric limit, the parameters of
the superpotential $W(\Phi)$ do not receive any (finite or infinite)
correction from Feynman diagrams involving the loops. In other
words, if the value of a superpotential parameter is fixed at the
classical level, it remains unchanged to all orders in perturbation
theory. This is known as the non-renormalization
theorem~\cite{ss|GRS79}.

This observation was realized as the key to solving the Higgs mass
problem of the standard model as follows: the radiative corrections
to the Higgs mass in the standard model are quadratically divergent
and admit the Planck scale as a natural cutoff if there is no new
physics upto that level. Since the Higgs mass is directly
proportional to the mass of the $W$-boson, the loop corrections
would push the $W$-boson mass to the Planck scale destabilizing the
standard model. On the other hand in the supersymmetric version of
the standard model (to be called MSSM), in the limit of exact
supersymmetry, there are no radiative corrections to any mass
parameter and therefore to the Higgs boson mass which can therefore 
be set once and for all at the tree level.
Thus if the world could be supersymmetric at
all energy scales, the weak scale stability problem would be easily
solved. However, since supersymmetry must be a broken symmetry, one
has to ensure that the terms in the Hamiltonian that break
supersymmetry do not spoil the non-renormalization theorem in a way
that infinities creep into the self mass corrections to the Higgs
boson.  This is precisely what happens if effective supersymmetry
breaking terms are ``soft" which means that they are of the following
type:
	\begin{enumerate}
\item $m^2_a \phi^\dagger_a \phi_a$, where 
$\phi$ is the bosonic component of the chiral superfield $\Phi_a$; 

\item $m\int d^2\theta \theta^2
\left(AW^{(3)}(\Phi)+BW^{(2)} (\Phi) \right)$, where $W^{(3)}(\Phi)$
and $W^{(2)}(\Phi)$ are the second and third order polynomials 
in the superpotential.

\item $\frac{1}{2} m_\lambda\lambda^T C^{-1} \lambda$, where
$\lambda$ is the gaugino field. 
	\end{enumerate}
It can be shown that the soft breaking terms only introduce finite
loop corrections to the parameters of the superpotential. Since all
the soft breaking terms require couplings with positive mass
dimension, the loop corrections to the Higgs mass will depend on
these masses and we must keep them less than a TeV so that the
weak scale remains stabilized.  This has the interesting implication
that superpartners of the known particles are accessible to the
ongoing and proposed collider experiments. For a recent survey of
the experimental situation, see Ref.~\cite{ss|susy96,howie,dawson}.

The mass dimensions associated with the soft breaking terms depend
on the particular way in which supersymmetry is broken. It is usually
assumed that supersymmetry is broken in a sector that involves
fields which do not have any quantum numbers under the standard
model group. This is called the hidden sector. The supersymmetry
breaking is then transmitted to the visible sector either via the
gravitational interactions \cite{ss|polonyi} or via the gauge
interactions of the standard model \cite{ss|dine} or via anomalous
U(1) ${\cal D}$-terms \cite{ss|binetruy}. In sec. 1.4, we discuss 
these different ways to break supersymmetry and their implications.

\subsection{The minimal supersymmetric standard model (MSSM)} 

Let us  now apply the discussions of the previous section to
constuct the supersymmetric extension of the standard model so that
the goal of stabilizing the Higgs mass is indeed realized in
practice. The superfields and their representation content are
given in Table I.


%
\begin{table}
\caption{The particle content of the supersymmetric standard
model. For matter and Higgs fields, we have shown the left-chiral  
fields only. The right-chiral fields will have a conjugate
representation under the gauge group.\label{ss!1}}
\begin{center}
\begin{tabular}{|c||c|}
\hline\hline
 Superfield &  gauge  transformation \\ \hline\hline
 Quarks $Q$ & $(3,2, {1\over 3})$\\
 Antiquarks $u^c$ &  $(3^*,1, - {4\over 3})$ \\
Antiquarks  $ d^c$ &  $(3^*, 1,\frac{2}{3})$\\
 Leptons $L$ & $(1, 2 -1)$ \\
 Antileptons  $e^c$ & $(1,1,+2)$ \\
Higgs Boson $\bf H_u$ & $(1, 2, +1)$ \\
Higgs Boson $\bf H_d$ & $(1, 2, -1)$ \\
Color Gauge Fields  $G_a$ & $(8, 1, 0)$ \\
Weak Gauge Fields  $W^{\pm}$, $Z$, $\gamma$ & $(1,3+1,0)$ \\
\hline\hline
\end{tabular}

\end{center}
\end{table}
 
First note that an important difference between the standard model
and its supersymmetric version apart from the presence of the   
superpartners is the presence of a second Higgs doublet. This is
required both to give masses to quarks and leptons as well as to
make the model anomaly free. The gauge interaction part of the model
is easily written down following the rules laid out in the previous 
section. In the weak eigenstate basis, weak interaction Lagrangian for the
quarks and leptons is exactly the same as in the standard model. As
far as the weak interactions of the squarks and the sleptons are
concerned, the generation mixing angles are very different from
those in the corresponding fermion sector due to supersymmetry
breaking. This has the phenomenological implication that the
gaugino-fermion-sfermion interaction changes generation leading to
potentially large flavor changing neutral current effects such as
$K^0$-$\bar {K}^0$ mixing, $\mu\to e\gamma$ decay etc unless the
sfermion masses of different generations are chosen to be very close
in mass.

Let us now proceed to a discussion of the superpotential of the
model.  It consists of two parts:
	\begin{eqnarray}
W= W_1 + W_2 \,,
	\end{eqnarray}
where
	\begin{eqnarray}
W_1 = h^{ij}_{\ell} e^c_i L_j {\bf H_d} + h^{ij}_d Q_id^c_j {\bf H_d} + 
h^{ij}_u Q_i u^c_j{\bf H_u} + \mu {\bf H_u H_d} 
\end{eqnarray}

\begin{eqnarray}
W_2 = \lambda_{ijk} L_iL_je^c_k +\lambda'_{ijk} L_iQ_jd^c_k 
+\lambda''_{ijk} u^c_id^c_jd^c_k
\label{ss.W}
	\end{eqnarray}
$i,j,k$ being generation indices. 
We first note that the terms in $W_1$ conserve baryon and lepton
number whereas those in $W_2$ do not. The latter are known as the
$R$-parity breaking terms where $R$-parity is defined as
	\begin{eqnarray}
R = (-1)^{3(B-L)+2S} \,,
	\end{eqnarray}
where $B$ and $L$ are the baryon and lepton numbers and
$S$ is the spin of the particle. It is
interesting to note that the $R$-parity symmetry defined above
assigns even $R$-parity to known particles of the standard model and
odd $R$-parity to their superpartners. This has the important
experimental implication that for theories that conserve $R$-parity,
the super-partners of the particles of the standard model must
always be produced in pairs and the lightest superpartner must be a
stable particle.  This is generally called the LSP. If the LSP turns
out to be neutral, it can be thought of as the dark matter particle
of the universe.

We now assume that some kind of supersymmetry breaking mechanism
introduces
splitting for the squarks and sleptons from the quarks and the
leptons. Usually, supersymmetry breaking can be expected to introduce 
trilinear scalar interactions amomg the sfermions as follows:
	\begin{eqnarray}
{\cal L}^{SB}=m_{3/2}[ A_{e, ab} \widetilde{e^c}_a\widetilde{L}_bH_d
+A_{d,ab}\widetilde{Q}_a 
H_d \widetilde{d^c}_b +A_{u, ab}\widetilde{Q}_a H_u
\widetilde{u^c}_b]\\ \nonumber
+B\mu m_{3/2} H_uH_d
 +\Sigma_{i= {\rm scalars}}
\mu^2_i\phi^{\dagger}_i\phi_i 
+\Sigma_a \frac{1}{2}M_a \lambda^T C^{-1} \lambda_a
\label{ss.SB}
	\end{eqnarray}
There will also be the corresponding terms involving the $R$-parity
breaking, which we omit here for simplicity.
   
As already announced this model solves the Higgs mass problem in the 
sense that if its tree level value is chosen to be of the order of the 
electroweak scale, any radiative correction to it will only induce 
terms of order $\sim \frac{f^2}{16\pi^2} M^2_{SUSY}$. By choosing the 
supersymmetry breaking scale in the TeV range, we can guarantee that
to all orders in perturbation theory the Higgs mass remains stable.

Constraints of supersymmetry breaking provide one prediction that can 
distinguish it from the nonsupersymmetric models- i.e. the mass of the
lightest Higgs boson. It can be shown that the
lightest higgs boson mass-square is going to be of order $\sim 
g^2v^2_{wk}$ (Ref.\cite{howie}). In fact denoting the vev's of the two Higgs
doublets as $<H^0_u>=v_u$ and $<H^0_d>=v_d$, one can write:
\begin{eqnarray}
m^2_{h}\simeq \frac{g^2+g'^2}{4}(v^2_d-v^2_u)
\end{eqnarray}
Defining $v_u/v_d=tan\beta$, we can rewrite the above light Higgs mass 
formula as $m^2_h= M^2_Z cos 2\beta$ which implies that the tree 
level mass of the lightest Higgs boson is less than the $Z$ mass. Once 
radiative corrections are taken into account\cite{howie}, $m_h$ increases
above the $M_Z$. However, it is now well established that in a large class
of supersymmetric models (which do not differ too much from the MSSM),
the Higgs mass is less than 150 GeV or so.

Another very interesting property of the MSSM is that 
electroweak symmetry breaking can be induced by radiative corrections.
As we will see below, in all the schemes for generating
soft supersymmetry breaking terms via a hidden sector, one generally gets 
positive (mass)$^2$'s for all scalar fields at the scale of SUSY breaking
as well as equal mass-squares.
In order to study the theory at the weak scale, one must extrapolate all
these parameters using the renormalization group equations. The degree of
extrapolation will of course depend on the strength of the gauge and the 
Yukawa couplings of the various fields. In particular, the $m^2_{H_u}$ 
will have a strong extrapolation proportional
to $\frac{h^2_t}{16\pi^2}$ since $H_u$ couples to the top quark. 
Since $h_t\simeq 1$, this can make $m^2_{H_u}(M_Z)< 0$,
leading to spontaneous breakdown of the electroweak symmetry. 
An approximate solution of the renormalization group
equations gives 
\begin{eqnarray}
m^2_{H_u}(M_Z)=m^2_{H_u}(\Lambda_{SUSY})-\frac{3h^2_tm^2_{\tilde{t}}}{16\pi^2}
ln\frac{\Lambda^2_{SUSY}}{M^2_Z}
\end{eqnarray}
This is a very attractive feature of supersymmetric theories.

\subsection{Why go beyond the MSSM ?}

Even though the MSSM solves two outstanding peoblems of the standard model,
i.e. the stabilization of the Higgs mass and the breaking of the electroweak
symmetry, it brings in a lot of undesirable consequences. They are:

(a) Presence of arbitrary baryon and lepton number violating couplings
i.e. the $\lambda$, $\lambda'$ and $\lambda''$ couplings described above.
In fact a combination of $\lambda'$ and $\lambda''$ couplings lead to
proton decay. Present lower limits on the proton lifetime then imply that
$\lambda'\lambda''\leq 10{-25}$ for squark masses of order of a TeV.
Recall that a very attractive feature of the standard model is the
automatic conservation of baryon and lepton number.
The presence of R-parity breaking terms\cite{ss|aul82} also makes it 
impossible to use the LSP as the Cold Dark 
Matter of the universe since it is not stable and will therefore decay away
in the very early moments of the universe. We will see that as we
proceed to discuss the various grand unified theories, keeping the
R-parity violating terms under control it will provide a major constraint on
model building.

(b) The different mixing matrices in the quark and squark sector leads to
arbitrary amount of flavor violation manifesting in such phenomena as
$K_L-K_S$ mass difference etc. Using present experimental information and 
the fact that the standard model more or less accounts for the observed
magnitude of these processes implies that there must be strong 
constraints on the mass splittings among squarks. Detailed calculations 
indicate\cite{ss|masiero} that one must have $\Delta 
m^2_{\tilde{q}}/m^2_{\tilde{q}}\leq 10^{-3}$ or so. Again recall that
this undoes another nice feature of the standard model.

(c) The presence of new couplings involving the super partners allows
for the existence of extra CP phases. In particular the presence of the
phase in the gluino mass leads to a large electric dipole moment of the 
neutron unless this phase is assumed to be suppressed by two to three
orders of magnitude\cite{garisto}. This is generally referred to in
the literature as the SUSY CP rpoblem. In addition, there is of course 
the famous strong CP problem which neither the standard model nor the MSSM
provide a solution to.

In order to cure these problems as well to understand the origin of the
soft SUSY breaking terms, one must seek new physics beyond the MSSM. Below,
we pursue two kinds of directions for new physics: one which analyses
schemes that generate soft breaking terms and a second one which leads to
automatic B and L conservation as well as solves the SUSY CP problem.
The second model also provides a solution to the strong CP problem 
without the need for an axion under certain circumstances. 

\subsection{Mechanisms for supersymmetry breaking}      

One of the major focus of research in supersymmetry is to understand the
mechanism for supersymmetry breaking. The usual strategy employed is to
assume that SUSY is broken in a hidden sector that does not involve any 
of the matter or forces of the standard model (or the visible sector)
and this SUSY breaking is transmitted to the visible sector via some
intermediary , to be called the messenger sector.

There are generally two ways to set up the hidden sector- a less 
ambitious one where one writes an effective Lagrangian (or superspotential)
in terms of a certain set of hidden sector fields that lead to 
supersymmetry breaking in the ground state and another more ambitious one
where the SUSY breaking arises from the dynamics of the hidden sector
interactions. For our purpose we will use the simpler schemes of the
first kind. As far as the messenger sector goes there are three possibilities
as already referred to earlier: (i) gravity mediated \cite{ss|polonyi};
(ii) gauge mediated \cite{ss|dine} and (iii) anomalous U(1) 
mediated\cite{ss|binetruy}. Below we give examples of each class.

\bigskip

\noindent{\it (i) Gravity mediated SUSY breaking}

\bigskip

 The scenario that uses gravity to transmit the supersymmetry breaking 
is one of the earliest hidden sector scenarios for SUSY breaking and
 forms much of the basis for the
discussion in current supersymmetry phenomenology.
In order to discuss these models one needs to know the supersgravity 
couplings to matter. This is given in the classic paper of Cremmer et 
al.\cite{ss|gira}. An essential feature of supergravity coupling is the
generalized kinetic energy term in gravity coupled theories called the Kahler
potential, $K$. We will denote this by $G$ and it is a hermitean operator 
which is a function of the matter fields in the theory and their complex 
conjugates. The effect of supergravity coupling in the matter and the gauge
sector of the theory is given in terms of $G$ and its derivatives as follows:
\begin{eqnarray}
L(z) = G_{zz^*}|\partial_{\mu}z|^2 + e^{-G}[G_zG_{z^*}G^{-1}_{zz^*}+3]
\end{eqnarray}
where $z$ is the bosonic component of a typical chiral field (e.g. we would
have $z\equiv \tilde{q}, \tilde{l}$ etc) 
and $G=~3ln(\frac{-K}{3})- ln|W(z)|^2$. A superscript implies derivative with
respect to that field.
The simplest choice for the Kahler potential $K$ is 
$K=-3e^{-\frac{|z|^2}{3M^2_{P\ell}}}$
that normalizes the kinetic energy term properly. Using this, one can the
effective potential for supergravity coupled theories to be:
\begin{eqnarray}
V(z,z^*)= e^{\frac{|z|^2}{M^2_{P\ell}}}[|W_z+\frac{z^*}{M^2_{P\ell}}W|^2
-\frac{3}{M^2_{P\ell}}|W|^2]+ D-terms
\end{eqnarray}
The gravitino mass is given interms of the Kahler potential as :
\begin{eqnarray}
m_{3/2}= M_{P\ell}e^{-G/2}
\end{eqnarray}

A popular scenario      
suggested by Polonyi is based upon the following hidden sector
consisting of a gauge singlet field, denoted by
$z$ and the superpotential $W_H$ given by:
        \begin{eqnarray}
W_H= \mu^2(z+\beta)
        \end{eqnarray}
where $\mu$ and $\beta$ are mass parameters to be fixed by various
physical considerations. It is clear that this superpotential leads to
an F-term that is always non-vanishing and theerfore breaks supersymmetry. 
Requiring the cosmological   
constant to vanish fixes $\beta=(2-\sqrt{3})M_{Pl}$. Given this  
potential and the choice of the Kahler potential as discussed earlier, 
supergravity calculus predicts 
a universal soft breaking parameters $m$ given by $m_0\sim
\mu^2/M_{Pl}$. Requiring $m_0$ to be in the TeV range implies that
$\mu\sim 10^{11}$ GeV. The complete potential to zeroth order in 
$M^{-1}_{P\ell}$ in this model is given by:
\begin{eqnarray}
V(\phi_a)= [ \Sigma_a |\frac{\partial W}{\partial \phi_a}|^2 + V_D]\\ \nonumber
+[m^2_0\Sigma_a \phi^*_a\phi_a + (A W^{(3)}+BW^{(2)}+h.c.)
\end{eqnarray}
where $W^{(3,2)}$ denote the dimension three and two terms in the 
superpotential respectively. The values of the parameters $A$ and $B$
at $M_{Pl}$ are related to each other in this example as $B=A-1$.
 The gaugino masses in these models arise
out of a separate term in the Lagrangian depending on a new function of the
hidden sector singlet fields, $z$:
\begin{eqnarray}
\int d^4x d^2\theta f(z) W^{\alpha}_{\lambda}W_{\lambda,\alpha}
\end{eqnarray}
If we choose $f(z)=\frac{z}{M_{P\ell}}$, then gaugino masses come out to 
be order $m_{3/2}\sim \frac{\mu^2}{M_{P\ell}}$ which is also of order
$m_0$ , i.e. the electroweak scale. Furthermore, in order to avoid
undesirable color and electric charge breaking by the SUSY models, one
must require that $m^2_0\geq 0$. 

 It is important to point out that the superHiggs  
mechanism operates at the Planck scale. Therefore all parameters
derived at the tree level of this model need to be extrapolated to
the electroweak scale.  So after the soft-breaking Lagrangian is 
extrapolated to the weak scale, it will look like:
        \begin{eqnarray}
{\cal L}^{SB} = m^2_{a}\phi^*_a\phi_a+ m\Sigma_{i,j,k} A_{ijk} 
\phi_i\phi_j\phi_k + \Sigma_{i,j} B_{ij}\phi_i\phi_j
\end{eqnarray}
These extrapolations depend among other things on the Yukawa couplings
of the model. As a result of this the universality of the various SUSY 
breaking terms is no more apparent at the electroweak scale. Moreover, since 
the top Yukawa coupling is now known to 
be of order one, its effect turns the mass-squared of the $H_u$ 
negative at the electroweak scale even starting from a positive value at the
Planck scale \cite{ss|alva}. This provides a natural mechanism for the 
breaking of electrweak symmetry adding to the attractiveness of 
supersymmetric models. In the lowest order approximation, one gets,
\begin{eqnarray}
m^2_{H_u}(M_Z)\sim m^2_{H_u}(M_{P\ell})-\frac{3h^2_t}{8\pi^2} 
ln(\frac{M_{P\ell}}{M_Z})(m^2_{H_u}+m^2_{\tilde{q}}+m^2_{u^c})|_{\mu=M_{P\ell}}
\end{eqnarray}

Before leaving this section it is worth pointing out that despite the 
simplicity and the attractiveness of this mechanism for SUSY breaking, there 
are several
serious problems that arise in the phenomenological study of the model that
has led to the exploration of other alternatives. For instance, the observed 
constraints on the flavor changing neutral currents\cite{ss|masiero} require 
that the squarks of the first and the second generation must be nearly 
degenerate, which is satisfied if one assumes the universality of the
spartner masses at the Planck scale. However this universality depends on 
the choice of the Kahler potential which is adhoc. 

Before we move on to the discussion of the alternative scenarios
for hidden sector, we point out an attractive choice for the Kahler
potential which leads naturally to the vanishing of the cosmological
constant unlike in the Polonyi case where we had to dial the large
cosmological constant to zero. The choice is $G=3 ln (S+S^{\dagger})$,
which as can easily be checked from the Eq. 11 to lead to $V=0$. This
is known as the no scale model\cite{nano} and usually emerges in the
case of string models\cite{witten1}. A complete and successful implementation
of this idea with the gravitino mass generated in a natural way in
higher orders is still not available.

\bigskip

\noindent{\it (ii) Gauge mediated SUSY breaking\cite{ss|dine}}

\bigskip

This mechanism for the SUSY breaking 
has recently been quite popular in the literature and involves
different hidden as well as messenger sectors. In particular, it proposes to
use the known gauge forces as the messengers of supersymmetry breaking.
As an example, consider a unified hidden messenger sector toy model of the
following kind, consisting of the fields $\Phi_{1,2}$ and $\bar{\Phi}_{1,2}$
which have the standard model gauge quantum numbers and a singlet field
$S$ and with the following superpotential:
\begin{eqnarray}
W= \lambda S (M^2_0-\bar{\Phi}_1\Phi_1) + 
M_1(\bar{\Phi}_1\Phi_2+\Phi_1\bar{\Phi}_2) +M_2\bar{\Phi}_1\Phi_1
\end{eqnarray}
The F-terms of this model are given by:
\begin{eqnarray}
F_S=\lambda(M^2_0-\bar{\Phi}_1\Phi_1)\\ \nonumber
F_{\Phi_2}= M_1\Phi_1;~ F_{\bar{\Phi}_2}= M_1\Phi_1 \\ \nonumber
F_{\Phi_1}= M_2\bar{\Phi}_1 +M_1\bar{\Phi}_2-\lambda S\bar{\Phi}_1
\end{eqnarray}
It is easy to see from the above equation that for $M_1\gg M_0, M_2$,
the minimum of the potential corresponds to all $\Phi$'s having zero vev
and $F_S=\lambda M^2_0$, thus breaking supersymmetry. The same 
superpotential responsible for SUSY breaking also transmits the
SUSY breaking information to the visible sector. While the spirit
of this model\cite{dutta} is similar to the original papers on the subject 
this unified construction is different and has its characteristic
predictions.

The SUSY breaking to the visible sector is transmitted via one and two 
loop diagrams. The gaugino masses arise from the one loop diagram
where a gaugino decomposes into the SUSY partners $\phi_1$ and
$\tilde{\phi}$ and the loop is completed as $\phi_1$ and $\bar{\phi}_1$
mix thru $F_S$ susy breaking term and the fermionic partners mix via
the mass term $M_2$.
The squark and slepton masses arise from the two loop diagram where
the squark-squark gauge boson -gauge boson coupling begins the first
loop and one of the gauge bosos couples to the two $\phi_1$'s and another
to the two $\bar{\phi}_1$'s which in turn mix via the F-terms for S to
complete the two loop diagram. This is only one typical diagram and
there are many more which contribute in the same order (see Martin, Ref. 20).
It is then easy to see that their magnitudes are given by:
\begin{eqnarray}
m_{\lambda}\simeq \frac{\alpha}{4\pi}\frac{<F_S>}{M_2}\\ \nonumber
m^2_{\tilde{q}}\simeq 
\left(\frac{\alpha}{4\pi}\right)^2\left(\frac{<F_S>}{M_2}\right)^2
\end{eqnarray}
The first point to notice is that the gaugino and squark masses are
roughly of the same order and requiring the squark masses to be around
100 GeV, we get for $F_S /M_2\simeq 100$ TeV. Of course, $<F_S>$ and
$M_2$ need not be of same order in which case the numerics will be 
different.
Another important point to note is that by choosing the quantum numbers 
of the messengers $\Phi_i$ appropriately, one can have widely differing 
spectra for the superpartners.

A distinguishing feature of this approach is that due to low scale for 
SUSY breaking, the gravitino mass is always in the milli-eV to kilo-eV
range and is therefore is always the LSP. Thus these models cannot lead to
a supersymmetric CDM.

The attractive property of these models is that they lead naturally to
near degeneracy of the squark and sleptons thus alleviating the FCNC
problem of the MSSM and have therefore been the focus of intense scrutiny 
during the past year\cite{GMSB}.

These class of models however suffer from the fact that the messenger 
sector is too adhoc .

\bigskip

\noindent{\it (iii) Anomalous U(1) mediated supersymmetry breaking}

\bigskip

These class of models owe their origin to the string models, which after 
compactification can often leave anomalous U(1) gauge groups\cite{dine1}. 
Since the 
original string model is anomaly free, the anomaly cancellation must take 
place via the Green-Schwarz mechanism as follows. Consider a U(1) gauge 
theory with a single chiral fermion that carries a U(1) quantum number.
This theory has an anomaly. Therefore, under a gauge transformation, the 
low energy Lagrangian is not invariant and changes as:
\begin{eqnarray}
L \rightarrow L+ \frac{\alpha}{4\pi} F\tilde{F}
\end{eqnarray}
where $F_{\mu\nu}=\partial_{\mu} A_{\nu}-\partial_{\nu}A_{\mu}$ and 
$\tilde{F}$ is the dual of $F_{\mu\nu}$. The last term is the anomaly 
term. To restore gauge invariance, we can add to the Lagrangian the 
Green-Schwarz term and rewrite the effective Lagrangian as
\begin{eqnarray}
L'=L+\frac{a}{M}F\tilde{F}
\end{eqnarray}
where under the gauge transformation $a\rightarrow a-M\alpha/4\pi$.
In order to obtain the supersymmetric version of the Green-Schwarz term,
we have to add a dilaton term to the axion $a$ to make a complex chiral 
superfield. Let us denote the dilaton field by $\phi$ and the complex
chiral field containing it as $S=\phi + i a$. The gauge invariant action
containing the $S$ and the gauge supersfield $V$ has terms of the following
form:
\begin{eqnarray}
A=\int d^4\theta ln(S+S^{\dagger}-V) + \int d^2\theta SW^{\alpha}W_{\alpha}
+ matter~ field~ parts
\end{eqnarray}
It is clear that in order to get a gauge field Lagrangian out of this, 
the dialton $S$ must have a vev with the identification that 
$<S>=g^{-2}$ and it is a fundamental unanswered question in superstring 
theory as to how this vev arises. If we assume that this vev has been 
generated, then, one can see that the first term in the Lagrangian when
expanded around the dilaton vev, leads to a term $\frac{1}{<2S>}\int 
d^4\theta
V$, which is nothing but a linear Fayet-Illiopoulos D-term. Combining this 
with other matter field terms with non-zero U(1) charge, one can then write
the D-term of the Lagrangian. As an example that can lead to realistic 
model building, we take two fields with equal and opposite 
U(1) charges $\pm 1$ in addition to the squark and slepton 
fields. The D-term can then be written as:
\begin{eqnarray} V_D=\frac{g^2}{2} 
(n^2_{q}|\tilde{Q}|^2 +n^2_{L}|\tilde{L}|^2 +|\phi_+|^2-|\phi_-|^2+\zeta)^2
\end{eqnarray}
This term when minimized does not break supersymmetry. However, if we add 
the superpotential a term of the form $W_{\phi}= m\phi_+\phi_-$, then 
there is
another term in low energy effective potential that leads to the combined
potential as:
\begin{eqnarray}
V= V_D + m^2(|\phi_+|^2 +|\phi_-|^2)
\end{eqnarray}
 The minimum of this potential corresponds to:
\begin{eqnarray}
<\phi_+> =0; <\phi_-> = (\zeta -\frac{m^2}{g^2})^{1/2}\simeq \epsilon 
M_{P\ell}: F_{\phi_+}= m M_{P\ell} \epsilon
\end{eqnarray}
where we have assumed that  $\zeta =\epsilon^2 M^2_{P\ell}$. This then 
leads to nonzero squark masses $m^2_{\tilde{Q}}\simeq n^2_{Q} m^2$. Thus 
supersymmetry is broken and superpartners pick up mass. In the simplest
model it turns out that the gaugino masses may be too low and one must
seek ways around this. However, the A and B-terms are also likely to be
small in this model and that may provide certain advantages. On the whole,
this approach has great potential for model building and has not been 
thoroughly 
exploited\cite{anom}- for instance, it can be used to solve the FCNC 
problems, SUSY CP problem, to study the fermion mass hierarchies etc. It is 
beyond the scope of this review to enter into those areas. One can expect 
to see activity in this area blossom.

\subsection{Supersymmetric Left-Right model}
	%
One of the attractive features of the
supersymmetric models is its ability to provide a candidate for the
cold dark matter of the universe. This however relies on the
theory obeying $R$-parity conservation (with $R\equiv (-1)^{3(B-L)+2S}$).
It is easy to check that particles of the standard model are even under
R whereas their superpartners are odd. The lightest superpartner is then
absolutely stable and can become the dark matter of the universe. In 
the MSSM, R-parity symmetry is not automatic and is 
achieved by imposing global baryon and lepton number conservation on
the theory as additional requirements.  First of all, this takes us
one step back from the non-supersymmetric standard model where the
conservation $B$ and $L$ arise automatically from the gauge symmetry
and the field content of the model. Secondly, there is a prevalent
lore supported by some calculations that in the presence of
nonperturbative gravitational effects such as black holes or worm
holes, any externally imposed global symmetry must be violated by
Planck suppressed operators \cite{ss|Gid88}. In this case, the
$R$-parity violating effects again become strong enough to cause
rapid decay of the lightest $R$-odd neutralino so that there
is no dark matter particle in the minimal supersymmetric standard model. 
It is therefore desirable to seek supersymmetric theories where, like the
standard model, $R$-parity conservation (hence Baryon and Lepton
number conservation) becomes automatic i.e. guaranteed by the field
content and gauge symmetry. It was realized in
mid-80's \cite{ss|Moh86} that such is the case in the supersymmetric
version of the left-right model that implements the see-saw
mechanism for neutrino masses. We briefly discuss this model in the section.

The gauge group for this model is $SU(2)_L\times SU(2)_R\times
U(1)_{B-L}\times SU(3)_c$. The chiral superfields denoting
left-handed and right-handed quark superfields are denoted by
$Q\equiv (u,d)$ and $Q^c\equiv (d^c, -u^c)$ respectively
and similarly the lepton superfields are given by $L\equiv (\nu, e)$
and $L^c\equiv (e^c, -\nu^c)$. The $Q$ and $L$ transform
as left-handed doublets with the obvious values for the $B-L$ and
the $Q^c$ and $L^c$ transform as the right-handed doublets
with opposite $B-L$ values.  The symmetry breaking is achieved by
the following set of Higgs superfields: $\phi_a(2,2,0,1)$ ($a=1,2$);
$\Delta (3, 1, +2, 1)$; $\bar{\Delta}(3,1,-2,1)$;
$\Delta^c(1,3,-2,1)$ 
and $\bar{\Delta^c} (1,3,+2,1)$. There are alternative Higgs multiplets
that can be employed to break the right handed $SU(2)$; however, this way of
breaking the $SU(2)_R\times U(1)_{B-L}$ symmetry automatically leads to
the see-saw mechanism for small neutrino masses\cite{ramond} as mentioned.

The superpotential for this theory has only a very limited number of 
terms and is given by (we have suppressed the generation index):

\begin{eqnarray}
W & = &
{\bf Y}^{(i)}_q Q^T \tau_2 \Phi_i \tau_2 Q^c +
{\bf Y}^{(i)}_l L^T \tau_2 \Phi_i \tau_2 L^c
\nonumber\\
  & +  & i ( {\bf f} L^T \tau_2 \Delta L + {\bf f}_c
{L^c}^T \tau_2 \Delta^c L^c)
\nonumber\\
  & +  & \mu_{\Delta} {\rm Tr} ( \Delta \bar{\Delta} ) +
\mu_{\Delta^c} {\rm Tr} ( \Delta^c \bar{\Delta}^c ) +
\mu_{ij} {\rm Tr} ( \tau_2 \Phi^T_i \tau_2 \Phi_j )
\nonumber\\
 & + & W_{\it NR}
\label{eq:superpot}
\end{eqnarray}
where $W_{\it NR}$ denotes non-renormalizable terms arising from
higher scale physics such as grand unified theories or Planck scale effects.
At this stage all couplings ${\bf Y}^{(i)}_{q,l}$, $\mu_{ij}$,
$\mu_{\Delta}$, $\mu_{\Delta^c}$, ${\bf f}$, ${\bf f}_c$ are
complex with $\mu_{ij}$, ${\bf f}$ and ${\bf f}_c$ being symmetric matrices.

The part of the supersymmetric action that arises from this
is given by
\begin{eqnarray}
{\cal S}_W = \int d^4 x \int d^2 \theta \, W +
\int d^4 x \int d^2 \bar{\theta} \, W^\dagger \, .   
\end{eqnarray}

It is clear from the above equation that this theory has no baryon
or lepton number violating terms. Since all other terms in the theory 
automatically conserve B and L, R-parity symmetry 
$(-1)^{3(B-L)+2S}$ is automatically conserved in the SUSYLR model. As a 
result, it allows for a 
dark matter particle provided the vacuum state of the theory respects 
R-parity. The desired vacuum state of the theory
which breaks parity and preserves R-parity corresponds to $<\Delta^c>\equiv 
v_R\neq 0$;
$<\bar{\Delta^c}>\neq 0$ and $<\tilde{\nu^c}>=0$. This reduces the gauge 
symmetry to that of the 
standard model which is then broken via the vev's
of the $\phi$ fields. These two together via 
the see-saw mechanism\cite{ramond} lead to a formula for neutrino masses
of the form $m_{\nu}\simeq \frac{m^2_f}{fv_R}$. Thus we see that the 
suppression of the $V+A$ currents at low energies and the smallness of 
the neutrino masses are intimately connected.

It turns out that left-right symmetry imposes rather strong
constraints on the ground state of this model.
 It was pointed out in 1993 \cite{ss|kuchi} that
if we take the minimal version of this
model, the ground state leaves the gauge symmetry unbroken. To break gauge
symmetry one must include singlets in the theory. However, in this case,
the ground state breaks electric charge unless R-parity is spontaneously
broken. Furthermore, R-parity can be spontaneously broken only if
$M_{W_R}\leq $ few TeV's. Thus the conclusion is that the renormalizable
version of the SUSYLR model with only singlets, $B-L=\pm 2$ triplets and
bidoublets can have a consistent electric charge conserving vacuum only if
the $W_R$ mass is in the TeV range and $R$-parity is
spontaneously broken. This conclusion can however be avoided either by making
some very minimal extensions of the model such as adding superfields 
$\delta (3,1,0,1)+\bar\delta(1, 3, 0, 1)$\cite{ss|kuchi2} or by adding
nonrenormalizable terms to the theory\cite{goran}.
Such extra fields often emerge if the model
is embedded into a grand unified theory or is a consequence of an underlying 
composite model.

In order to get a R-parity conserving vacuum (as would be needed if we 
want the LSP to play the role of the cold dark matter) without 
introducing the
extra fields mentioned earlier, one must add the non-renormalizable terms.
In this case, the doubly charged Higgs bosons and Higgsinos become very
light unless the $W_R$ scale is above $10^{10}$ GeV or so\cite{chacko}
(and Aulakh et al. Ref.\cite{goran}). This implies that the neutrino masses
must be in the eV range, as would be required if they have to play the
role of the hot dark matter. Thus an interesting connection between the
cold and hot dark matter emerges in this model in a natural manner.

This model solves two other problems of the MSSM: (i) one is the 
SUSY CP problem and (ii) the other is the strong CP problem when the
$W_R$ scale is low. To see how this happens, let us define the
the transformation of the fields under left-right
symmetry as follows and observe the resulting constraints on the 
parameters of the model. 

\begin{eqnarray}
Q             & \leftrightarrow  &  {Q^c}^* \nonumber\\
L             & \leftrightarrow  &  {L^c}^* \nonumber\\
\Phi_i        & \leftrightarrow  &  {\Phi_i}^\dagger \nonumber\\
\Delta        & \leftrightarrow  &  {\Delta^c}^\dagger \nonumber\\
\bar{\Delta}  & \leftrightarrow  &  {\bar{\Delta}}^{c\dagger} \nonumber\\
\theta        & \leftrightarrow  &  \bar{\theta} \nonumber\\
{\tilde{W}}_{SU(2)_L} & \leftrightarrow  & {\tilde{W}}^*_{SU(2)_R} 
\nonumber\\
{\tilde{W}}_{B-L,SU(3)_C} &
         \leftrightarrow  & {\tilde{W}}^*_{B-L,SU(3)_C}
 \label{eq:lrdef}
\end{eqnarray}
   
Note that this corresponds to the usual definition
$Q_L \leftrightarrow Q_R$, etc. To study its implications on the 
parameters of the theory, let us write down the most general soft 
supersymmetry terms allowed by the symmetry of the model (which make the 
theory realistic).

\begin{eqnarray}
{\cal L}_{\rm soft} & = & \int d^4 \theta \sum_i m^2_i \phi_i^\dagger \phi_i
                      + \int d^2 \theta \, \theta^2 \sum_i A_i W_i 
     + \int d^2 \bar{\theta} \, {\bar{\theta}}^2 \sum_i A_i^* W_i^\dagger
                        \nonumber\\
     & + & \int d^2 \theta \, \theta^2 \sum_p m_{\lambda_p}
                 {\tilde{W}}_p {\tilde{W}}_p +
           \int d^2 \bar{\theta} \, {\bar{\theta}}^2 \sum_p m_{\lambda_p}^*
                 {{\tilde{W}}_p}^* {{\tilde{W}}_p}^* \, .
\label{eq:soft}
\end{eqnarray}
   
In Eq. \ref{eq:soft},  ${\tilde{W}}_p$ denotes the gauge-covariant
chiral superfield that contains the $F_{\mu\nu}$-type terms with
the subscript going over the gauge groups of the theory including
SU$(3)_c$. $W_i$ denotes the various terms in the superpotential, 
with all superfields replaced by their scalar components and
with coupling matrices which are not identical to those in $W$.
Eq. \ref{eq:soft} gives the most general set of soft breaking terms
for this model.
 
With the above definition of L-R symmetry, it is easy to check that
   
\begin{eqnarray}
{\bf Y}^{(i)}_{q,l} & = & {{\bf Y}^{(i)}_{q,l}}^\dagger \nonumber\\
\mu_{ij} & = & \mu_{ij}^* \nonumber\\
\mu_\Delta & = & \mu_{\Delta^c}^* \nonumber\\
{\bf f} & = & {\bf f}_c^* \nonumber\\
m_{\lambda_{SU(2)_L}} & = & m_{\lambda_{SU(2)_R}}^* \nonumber\\
m_{\lambda_{B-L,SU(3)_C}} &
          = & m_{\lambda_{B-L,SU(3)_C}}^*\nonumber\\
A_i & = & A^\dagger_i,
\label{eq:rels}  
\end{eqnarray}

Note that the phase of the gluino mass term is zero due to the constraint of
parity symmetry. As a result
the one loop contribution to the electric dipole moment of neutron from
this source vanishes in the lowest order\cite{rasin2}. The higher order loop
contributions that emerge after left-right symmetry breaking can be
shown to be small, thus solving the SUSYCP problem. 
Further more, since the constraints of left-right 
symmetry imply that the quark Yukawa matrices are hermitean, if the
vaccum expectation values of the $<\phi>$ fields are real, then
the $\Theta$ parameter of QCD vanishes naturally at the tree level. This
then provides a solution to the strong CP problem. It however
turns out that to keep the one loop finite contributions to the 
$\Theta$ less than $10^{-9}$, the $W_R$ scale must be in the TeV 
range\cite{kmr}. Such models generally predict the electric dipole moment
of neutron of order $10^{-26}$ ecm\cite{posp} which can be probed
in the next round of neutron dipole moment searches. 

The phenomenology of this model has been extensively
studied \cite{ss|FFK91} in recent papers and we do not go into
them here. A particularly interesting phenomenological prediction
of the model is the existence of the light doubly charged Higgs bosons 
and the corresponding Higgsinos.

\section{Unification of Couplings}

Soon after the discovery of the standard model, it became clear that
embedding the model into higher local symmetries may lead to two very
distinct conceptual advantages: (i) they may provide quark lepton unification
\cite{pati,georgi} providing a unified understanding of the apriori
separate interactions of the two different types of matter and (ii) they
can lead to description of different forces in terms of a single gauge 
coupling constant\cite{georgi,quinn}. How actually the unification of 
gauge couplings occurs was discussed in a seminal paper by 
Georgi, Quinn and Weinberg\cite{quinn}. They used the already
known fact that the coupling parameters in a theory depend on the mass scale
and showed that the gauge couplings of the standard model can indeed unify
at a very high scale of order $10^{15}$ GeV or so. Although 
this scale might appear too far removed from the energy 
scales of interest in particle physics then, it was actually 
a blessing in disguise since in GUT theories, obliteration of the 
quark-lepton distinction manifests itself in the form of baryon instability
such as proton decay and the rate of proton decay is inversely 
proportional to the 4th power of the grand unification scale and only for 
scales near $10^{15}$ GeV or so, already known lower limits on proton 
life times could be reconciled with theory.
This provided a new impetus for new experimental searches for proton decay.
The minimal grand unification model based on the
SU(5) group suggested by Georgi and Glashow made very precise prediction for
the proton lifetime of $\tau_p$ between $1.6\times 10^{30}$ yrs. to
$2.5\times 10^{28}$ yrs. Attempts to
observe proton decay at this level failed ruling out the simple minimal
nonsupersymmetric SU(5) model. In fact the situation was worse since
the minimal non-supersymmetric SU(5) also predicted a value for 
$sin^2\theta_W$ which is much lower than the experimentally observed one.

A revival of interest in the idea of grand unification occurred after the
ideas of supersymmetry became part of phenomenology of particle physics 
in the early 80's. Two points were realized that led to this. First is that
a theoretical understanding of the large hierarchy between the weak scale 
and the GUT scale was possible only within the framework of supersymmetry
as discussed in the first chapter. Secondly, on a more phenomenological 
level, measured values of $sin^2\theta_W$ from the accelerators coupled with
the observed values for $\alpha_{strong}$ and $\alpha_{em}$ could be 
reconciled with the unification of gauge couplings only if the superpartners
were included in the evolution of the gauge couplings and the supersymmetry
breaking scale was assumed to be near the weak scale, which was independently
motivated anyway\cite{many}.

It should be however made clear that supersymmetry is not the only
well motivated beyond standard model physics that leads to coupling
constant unification consistent with the measured value of $sin^2\theta_W$.
If the neutrinos have masses in the micro-milli-eV range, then the see-saw
mechanism\cite{ramond} given by the formula
\begin{eqnarray}
m_{\nu_i}\simeq \frac{m^2_{u_i}}{M_{B-L}}
\end{eqnarray}
implies that the $M_{B-L}$ scale is around $10^{11}$ GeV or so. It was 
shown in
the early 80's\cite{parida} that coupling constant unification can take place
without any need for supersymmetry if it is assumed that above the $M_{B-L}$
the gauge symmetry becomes $SU(2)_L\times SU(2)_R\times U(1)_{B-L}\times 
SU(3)_c$ or $SU(2)_L\times SU(2)_R\times SU(4)_c$. Since the subject of 
these lectures is supersymmetric grand unification, I will not discuss these
models here. Let us now proceed to discuss the unification of gauge 
couplings in supersymmetric models.
 
\subsection{Unification of Gauge Couplings (UGC)}

Key ingredients in this discussion are the renormalization group 
equations (RGE)
for the gauge coupling parameters. Suppose we want to evolve a coupling
parameter between the scales $M_1$ and $M_2$ (i.e. $M_1\leq \mu\leq M_2$)
corresponding to the two scales of physics. Then the RGE's depend on the
gauge symmetry and the field content at $\mu= M_1$. The one loop 
evolution equations for the gauge couplings (define $\alpha_i\equiv 
\frac{g^2_i}{4\pi}$) are:
 \begin{eqnarray}
\frac{d\alpha_i}{dt}=\frac{1}{2\pi} b_i \alpha^2_i
\end{eqnarray}
where $t=ln \mu$. The coefficient $b_i$ receives contributions from the 
gauge part and the matter including Higgs field part. In general,
\begin{eqnarray}
b_i=-3 C_2(G)+T(R_1)d(R_2)
\end{eqnarray}
where $C_2(R)=\Sigma_aR_aR_a$ and $T(R)\delta^{ab}= Tr(R_aR_b)$. $R_a$
are the generators of the gauge group under consideration. The following
group theoretical relations are helpful in making actual calculations:
\begin{eqnarray}
C_2(R)d(R)=T(R) r
\end{eqnarray}
where $d(R)$ is the dimension of the irreducible representation and $r$
is the rank of the group (the number of diagonal generators).

An important point to note is that since at the GUT scale one imagines
that the symmetry group merges into one GUT group, all the low energy
generators must be normalized the same way. What this means is that
if $\Theta_a$ are the generators of the groups at low energy, one must
satisfy the condition that $Tr (\Theta_a \Theta_b)= 2\delta_{ab}$.
If we sum over the fermions of the same generation, we easily see that
this condition is satisfied for the $SU(2)_L$ and the $SU(3)_c$ groups.
On the other hand, for the hypercharge generator, one must write
$\Theta_Y=\sqrt{\frac{3}{5}}\frac{Y}{2}$ to satisfy the correct normalization
condition. This must therefore be used in evaluating the $b_1$.

One can calculate the $b_i$ for the MSSM and they
are $b_3=-3$, $b_2=+1$ and $b_1=+33/5$ where the subscript $i$ denotes
the $SU(i)$ group (for $i> 1$) and we have assumed three generations of 
fermions. The gauge coupling evolution equations can then be written as:
\begin{eqnarray}
2\pi\frac{d\alpha^{-1}_1}{dt}=-\frac{33}{5}\\ \nonumber
2\pi\frac{d\alpha^{-1}_2}{dt}=-1 \\ \nonumber
2\pi\frac{d\alpha^{-1}_3}{dt}=3 
\end{eqnarray}
The solutions to these equations are:
\begin{eqnarray}
\alpha^{-1}_1(M_Z)=\alpha^{-1}_U+\frac{33}{10\pi}ln\frac{M_U}{M_Z}\\ 
\nonumber
\alpha^{-1}_2(M_Z)=\alpha^{-1}_U+\frac{1}{2\pi}ln\frac{M_U}{M_Z}\\ \nonumber 
\alpha^{-1}_1(M_Z)=\alpha^{-1}_U-\frac{3}{2\pi}ln\frac{M_U}{M_Z}\\ 
\end{eqnarray}
If these three equations which have only two free parameters hold then
coupling constant unification occurs. These equations lead to the 
consistency equation\cite{brahma}:
\begin{eqnarray}
\Delta \alpha\equiv 
5\alpha^{-1}_1(M_Z)-12\alpha^{-1}_2(M_Z)+7\alpha^{-1}_3(M_Z)=0
\end{eqnarray}
Using the values of the three gauge coupling
parameters measured at the $M_Z$ scale, i.e.
\begin{eqnarray}
\alpha^{-1}_1(M_Z)=58.97\pm .05\\ \nonumber
\alpha^{-1}_2(M_Z)=29.61\pm .05\\ \nonumber
\alpha^{-1}_3(M_Z)=8.47\pm .22
\end{eqnarray}
(where we have taken for the strong coupling constant the global average
given in Ref.\cite{schmelling}), we find that $\Delta\alpha= -1\pm 2$.
Thus we see that grand unification of couplings occurs
 in the one loop approximation. Again subtracting any two of the
above evolution equations, we find the unification scale to be $M_U\sim 
10^{16}$ GeV and $\alpha^{-1}_U\simeq 24$.
 There are of course two loop effects, corrections 
arising from the fact that all particle masses 
may not be degenerate and turn on as a Theta function in the evolution 
equations etc.\cite{mar}.
Another point worth noting is that while the value of $\alpha_{1,2}(M_Z)$
are quite accurately known, the same is not the case for the strong 
coupling constant and in fact detailed two loop calculations and the
MSSM threshold corrections reveal that\cite{bagger} if the effective
MSSM scale ($T_{SUSY}$) is less than $M_Z$, one needs $\alpha_s(M_Z)> .121$
to achieve unification. Thus indications of a smaller value for the 
QCD coupling would indicate more subtle aspects to coupling constant 
unification such as perhaps intermediate scales\cite{brahma} or new
particles etc.

To better appreciate the degree of unification in supersymmetric models, 
let us 
compare this with the evolution of couplings in the standard model.
The values of $b_i$ for this case  
are $b_1=\frac{41}{10}$, $b_2=-\frac{19}{6}$ and
$b_3=-7$. The gauge coupling unification in this case would require that
\begin{eqnarray}
\Delta\alpha_{sm}=\frac{6}{23}\alpha^{-1}_3(M_Z)-
\frac{309}{2507}\alpha^{-1}_2(M_Z)+\frac{15}{109}\alpha^{-1}_1(M_Z)=0
\end{eqnarray}
Using experimental inputs as before, it easy to check that $\Delta\alpha_{sm}
=6.5\pm .2$, which is away from zero by many sigma's.

\subsection{Gauge coupling unification with intermediate scales before
grand unification}

An important aspect of grand unification is the possibility that there
are intermediate symmetries before the grand unification symmetry is
realized. For instance a very well motivated example is the presence
of the gauge group $SU(2)_L\times SU(2)_R\times U(1)_{B-L}\times SU(3)_c$
before the gauge symmetry enlarges to the SO(10) group. So it is 
important to discuss how the evolution equation equations are modified
in such a situation.

Suppose that at the scale $M_I$, the gauge symmetry enlarges. To take 
this into account, we need to follow the following steps:

(i) If the smaller group $G_1$ gets embedded into a single bigger group
$G_2$ at $M_I$, then at the one loop level, we simply impose the matching 
condition:
\begin{eqnarray}
g_1(M_I)=g_2(M_I)
\end{eqnarray}

(ii) On the other hand if the generators of the low scale symmetry arise as
linear combinations of the generators of different high scale groups as
follows:
\begin{eqnarray}
\lambda_1=\Sigma_bp_b\theta_b
\end{eqnarray}
then the coupling matching condition is:
\begin{eqnarray}
\frac{1}{g^2_1(M_I)}=\Sigma_b \frac{p^2_b}{g^2_b(M_I)}
\end{eqnarray}

One can prove this as follows: for simplicity let us consider only the
case where $G_2=\Pi_b U(1)_b$ which at $M_I$ breaks down to a single
$U(1)$. Let this breaking occur via the vev of a single Higgs field
$\phi$ with
charges $(q_1,q_2,....)$ under $U(1)$. The unbroken generator is given by:
\begin{eqnarray}
Q=\Sigma p_aQ_a
\end{eqnarray}
with $\Sigma_cp_cq_c=0$.
The gauge field mass matrix after Higgs mechanism can be written as
\begin{eqnarray}
M^2_{ab}= g_ag_bq_aq_b<\phi>^2
\end{eqnarray}
This mass matrix has the following massless eigenstate which can be
identified with the unbroken U(1) gauge field:
\begin{eqnarray}
A_{\mu}=\frac{1}{(\Sigma_b \frac{p^2_b}{g^2_b})^{1/2}}\Sigma\frac{p_b}{g_b}
A_{\mu,b}\equiv N\Sigma_b\frac{p^2_b}{g^2_b}A_{\mu,b}
\end{eqnarray}
 To find the effective gauge coupling, we write
\begin{eqnarray}
L\sim \Sigma g_bQ_bA_{\mu,b}\\ \nonumber
=\Sigma_bg_b(p_b Q+...)N(\frac{p_b}{g_b}A_{\mu}+...)
\end{eqnarray}
Collecting the coefficient of $A_{\mu}$ and using the normalization 
condition $\Sigma p^2_b=1$, we get the result we wanted to prove (i.e. Eq.).

Let us apply this to the situation where $SU(2)_R\times U(1)_{B-L}$ is
broken down to $U(1)_Y$. In that case:
\begin{eqnarray}
\frac{Y}{2}= I_{3,R}+\frac{B-L}{2}
\end{eqnarray}
The normalized generators are $I_Y=\left(\frac{3}{5}\right)^{1/2}\frac{Y}{2}$
and $I_{B-L}=\left(\frac{3}{2}\right)^{1/2}\frac{B-L}{2}$. Using them, one
finds that 
\begin{eqnarray}
I_{Y}=\sqrt{\frac{3}{5}}I_{3R}+\sqrt{\frac{2}{5}}I_{B-L}
\end{eqnarray}
This implies that the matching of coupling constant at the scale where
the left-right symmetry begins to manifest itself is given by:
\begin{eqnarray}
\alpha^{-1}_Y=\frac{3}{5}\alpha^{-1}_{2R}+\frac{2}{5}\alpha^{-1}_{B-L}
\end{eqnarray}

\bigskip

\noindent{\it  Application to SO(10) GUT and possibility of low $W_R$ scale}

\bigskip

Let us apply this to the SO(10) model to see under what conditions a low
$W_R$ mass can be consistent with coupling constant unification. Let us 
first derive the evolution equations for the couplings in 
$SO(10)$ model with an intermediate $W_R$ scale.
For the $\alpha_2$ and $\alpha_3$, the evolution equations are 
starightforward and given by:
\begin{eqnarray}
\alpha^{-1}_i(M_Z)=\alpha^{-1}_U-\frac{b_i}{2\pi}ln\frac{M_R}{M_Z}-
\frac{b'_i}{2\pi}ln\frac{M_U}{M_R}
\end{eqnarray}
where $i=2,3$ and $b'_i$ receives contributions from all particles at and 
below the scale $M_R$. We assume that there are no other particles between 
$M_R$ and $M_U$ other than those included in $b_i$. Turning to $\alpha_1$,
we have to use the matching formula for the couplings derived above.
Using that, we find that
\begin{eqnarray}
\alpha^{-1}_1(M_Z)=\alpha^{-1}_1(M_R)-\frac{b_1}{2\pi}ln\frac{M_R}{M_Z}
\end{eqnarray}
Using the matching formula derived above, and evolving the $\alpha_{2R,B-L}$
between $M_R$ and $M_U$, we find that
\begin{eqnarray}
\alpha^{-1}_1(M_Z)=\alpha^{-1}_U-\frac{b_1}{2\pi}ln\frac{M_R}{M_Z}
-(\frac{3}{5}\frac{b'_{2R}}{2\pi}
-\frac{2}{5}\frac{b'_{BL}}{2\pi})ln\frac{M_U}{M_R}
\end{eqnarray}
In the discussion that follows let us denote $3/5 b'_{2R}+2/5 b'_{BL}\equiv
b'_1$. We then see that a sufficient condition for the intermediate scale
to exist is that we must have
\begin{eqnarray}
\Delta\alpha_{SUSY}\equiv 5b'_{1}-12 b'_{2L}+7b'_{3}=0
\end{eqnarray}
In fact if this condition is satisfied, at the one loop level, one can even
have a $W_R$ mass in the TeV range and still have coupling constant 
unification. As an example of such a theory, consider the following spectrum
of particles above $M_R$: a color octet, a pair of $SU(2)_R$ triplets
with $B-L=\pm 2$, two bidoublets $\phi(2,2,0)$ and a left-handed triplet. 
 The corresponding b-coefficients above $M_R$ are given by:
\begin{eqnarray}
b'_3=0; b'_{2L}=4; b'_{2R}=6~~ and~~ b'_{BL}=15
\end{eqnarray}
This theory satisfies the condition that $\Delta \alpha_{SUSY}=0$ and can
support a low $W_R$ theory.

\subsection{Yukawa unification}

Another extension of the idea of gauge coupling unification is to demand
the unification of Yukawa coupling parameters and study its implications
and predictions. This however is a much more model dependent conjecture 
than the UGC\cite{chan}.  One may of course demand partial Yukawa unification
instead of a complete one between all three generations. As we will see 
in the next chapter, most guand unification models tend to imply partial
Yukawa unification of type:
\begin{eqnarray}
h_b(M_U)=h_{\tau}(M_U)
\end{eqnarray}
To discuss the implications of this hypothesis, we need the 
renormalization group evolution of these couplings down to the 
weak scale. For this purpose, we need the R.G.E's for these couplings:
\begin{eqnarray}
2\pi \frac{dln Y_b}{dt}=6 Y_b+Y_t-\frac{7}{15}\alpha_1-\frac{16}{3}\alpha_3
-3\alpha_2\\ \nonumber
2\pi \frac{dln Y_{\tau}}{dt}=4Y_{\tau}-\frac{9}{5}\alpha_1-3\alpha_2\\
\nonumber
2\pi\frac{dln Y_t}{dt}=6Y_t+Y_b-\frac{13}{15}\alpha_1-\frac{16}{3}\alpha_3
-3\alpha_2
\end{eqnarray}
where we have defined $Y_i\equiv h^2_i/4\pi$. Subtracting the first two
equations in Eq. (59) and defining $R_{b/\tau}\equiv \frac{Y_b}{Y_{\tau}}$,
one finds that
\begin{eqnarray}
2\pi\frac{d}{dt}(R_{b/\tau})\simeq (Y_t-\frac{16}{3}\alpha_3)
\end{eqnarray}
Solving this equation using the Yukawa unification condition, we find that
\begin{eqnarray}
\frac{m_b}{m_{\tau}}(M_Z) = (R_{b/\tau}(M_Z))^{1/2}= A^{-1/2}_t\left(
\frac{\alpha_3(M_Z)}{\alpha_3(M_U)}\right)^{8/9}
\end{eqnarray}
where $A_t=e^{\frac{1}{2\pi}\int^{M_U}_{M_Z} Y_t dt}$.
Using the value of $\alpha_U$ from MSSM grand unification, we find that
$m_b/m_{\tau}(M_Z)\simeq 2.5 A^{-1/2}_t$. The observed value of 
$m_b/m_{\tau}(M_Z)\simeq 1.62$. So it is clear that a significant
contribution from the running of the top Yukawa is needed and this is
lucky since the top quark is now known to have mass of $\sim 175 $ GeV
implying an $h_t\simeq 1$. One way to estimate the $A_t$ is to assume
that $h_t(M_U)=3$, which case one has\cite{barger} $A^{-1/2}\simeq .85$
making $m_b/m_{\tau}$ closer to observations.

 It is worth pointing out that both the top Yukawa as well as the gauge 
contributions depend on whether there exist an intermediate scale. The
modified formula in that case is
\begin{eqnarray}
\frac{m_b}{m_{\tau}}(M_Z)= 
(A^{ZI}_tA^{IU}_t)^{-1/2}
\left(\frac{\alpha_3(M_Z)}{\alpha_3(M_I)}\right)^{8/9}\left(
\frac{\alpha_3(M_I)}{\alpha_3(M_U)}\right)^{8/3b'_3}
\end{eqnarray}

\bigskip

\noindent{\it  (A) Top Yuakawa coupling and its infrared fixed point}
\bigskip

It was noted by Hill and Pendleton and Ross\cite{hill}
for large Yukawa couplings, $h_t$, regardless of how large the asymptotic 
value is, the low energy value determined by the RGE's is a fixed
value and one can therefore use this observation to predict the top 
quark mass. To see this in detail, let us define a parameter $\rho_t
=Y_t/\alpha_3$. Using the RGE's for $Y_t$ and $\alpha_3$, we can then
write
\begin{eqnarray}
\alpha_3\frac{d\rho_t}{d\alpha_3}=-2\rho_t(\rho_t-\frac{7}{18})
\end{eqnarray}
The solution of this equation is
\begin{eqnarray}
\rho_t(\alpha_3)=\frac{7/8}{1-(1-\frac{7}{18\rho_{t0}})
(\frac{\alpha_3}{\alpha_{30}})^{-7/9}}
\end{eqnarray}
where $\rho_{t0}=\rho_t(\Lambda)$ and $\alpha_{30}=\alpha_{3}(\Lambda)$.
As we move to smaller $\mu$'s, $\alpha_3$ increases and as 
$\alpha_3\rightarrow $infinity, $\rho_t\rightarrow 7/18$. This leads to
$m_t=129 sin\beta$ GeV which is much smaller than the observed value.
Does this mean that this idea does not work ? The answer is no because,
strictly, at $m_t(m_t)$, the $\alpha_3$ is far from being infinity.
A more sensible thing to do is to use the RGE's for $Y_t$ and assume that
at $\Lambda\gg M_Z$, $Y_t\gg \alpha_3$ so that for very large $\mu$,
we have 
\begin{eqnarray}
\frac{dY_t}{dt}\simeq \frac{3Y^2_t}{\pi}
\end{eqnarray}
As a result, as we move down from $\Lambda$, first $Y_t$ will decrease
till it bocomes comparable to $\alpha_3$ after which, it will settle down
to the value $6Y_t=16/3 \alpha_3$ for which $Y_t$ stops running. This
leads to a prediction of $m_t\simeq 196 sin\beta$ GeV, which is more
consistent with observations. Note incidentally that if we applied the same
arguments to the standard model, we would obtain $m_t\simeq 278$ GeV,
which is much too large. Could this be an indication that supersymmetry
is the right way to go in understanding the top quark mass ?

Finally, we wish to very briefly mention that one could have demanded
complete Yukawa unification as is predicted by simple SO(10) 
models\cite{chan}:
\begin{eqnarray}
h_t(M_U)=h_b(M_U)=h_{\tau}(M_U)\equiv h_U
\end{eqnarray}
Extrapolating this relation to $M_Z$ one could obtain $m_t,m_b,m_{\tau}$
in terms of only two parameters $h_U$ and $tan\beta$. This would be a way to
also predict $m_t$. This is therefore an attractive idea. But getting the
electrweak symmetry breaking in this scenario is very hard since both
$m^2_{H_u}$ and $m^2_{H_d}$ run parallel to each other except for a minor
difference arising from the $U(1)_Y$ effects. One therefore has to make 
additional assumptions to understand the electrweak symmetry breaking
out of radiative corrections.

\bigskip

\section{Supersymmetric SU(5)}

\bigskip

The simplest supersymmetric grand unification model is based on the
simple group SU(5)\cite{dimo} and it embodies many of the unification
ideas discussed in the previous chapter. It is assumed that at the GUT 
scale $M_U$, SU(5) gauge symmetry breaks down to MSSM as follows:
\begin{eqnarray}
SU(5) \rightarrow SU(3)_c \times SU(2)_L \times U(1)_Y
\end{eqnarray}
The unification ideas of the previous section tell us that the single
gauge coupling at the GUT scale branches down to the three couplings of the
standard model. 

\bigskip

\subsection{Particle assignment and symmatry breaking}

\bigskip

To discuss further properties of the model, we discuss the
assignment of the matter fields as well as the Higgs superfields to the 
simplest representations necessary. The matter fields are assigned to the
$\bar{5}\equiv \bar{F}$ and $10\equiv 10$ dimensional representations 
whereas the Higgs fields are assigned to $\Phi\equiv 45$, $H\equiv {5}$ and
$\bar{H}\equiv \bar{5}$ representations.

\noindent{\it \underline{Matter Superfields:}}
\begin{equation}
\bar{F} =\left(\begin{array}{c} 
d^c_1\\ d^c_2\\ d^c_3 \\e^-\\ \nu\end{array}\right)\\ \nonumber
  ; T \{ 10 \}=\left(\begin{array}{ccccc}
0 & u^c_3 & -u^c_2 & u_1 & d_1\\
-u^c_3 & 0 & u^c_1 &u_2 & d_2 \\
u^c_2 & -u^c_1 & 0 & u_3 & d_3 \\
-u_1 & -u_2 & u_3 & 0 & e^+\\
-d_1 & -d_2 & -d_3 & -e^+ & 0 \end{array} \right)
\end{equation}
In the following discussion, we will choose the group indices as 
$\alpha, \beta$ for SU(5);
(e.g.$ H^\alpha, \bar{H}_\alpha, \bar{F}_{\alpha}
T^{\alpha\beta}= -T^{\beta\alpha}$ );
$i,j,k..$ will be used for $ SU(3)_c $ indices and 
$p,q$ for $ SU(2)_L$ indices.

To discuss symmetry breaking and other dynamical aspects of the model, we 
choose the superpotential to be: 
\begin{equation}
W = W_Y + W_G + W_h + W\prime
\end{equation}
where
\begin{eqnarray}
W_Y = h_u^{ab} \epsilon_{\alpha\beta\gamma\delta\sigma} T_a^\alpha\beta
T_b^{\gamma\delta} H^\sigma + h_d^ab T^{\alpha\beta} \bar{F}_\alpha
\bar{H}_\beta
\end{eqnarray}
($a,b$ are generation indices). This part of the superpotential is 
resposible for giving mass to the fermions.

\begin{eqnarray}
W_{G}= zTr \Phi+ x Tr{\Phi}^2 + y Tr{\Phi}^3 + \lambda_1 (H\Phi\bar{H} +
MH\bar{H})
\end{eqnarray}
This part of the superpotential is responsible for symmetry breaking
and getting light Higgs doublets below $M_U$. Note that although 
$Tr\Phi=0$ the z-term added as a Lagrange multiplier to enforce this 
constraint during potential minimization. Of the rest of the
superpotential
$W_h $ is the Hidden sector supetrpotential responsible for supersymmetry
breaking and 
$W\prime$ denotes the R-parity breaking terms which will be discussed
later.
We are looking for the following symmetry breaking chain:
\begin{equation}
SU(5) \times SUSY \rightarrow <\Phi>\neq 0 \rightarrow G_{std}\times SUSY
\end{equation}
To sudy this we have to use $W_G$ and calculate the relevant F-terms and 
set them to zero to maintain supersymmetry down to the weak scale.
\begin{eqnarray}
F^{\alpha}_{\Phi,\beta}=z\delta^{\alpha}_{\beta}+2x\Phi^{\alpha}_{\beta}
+3y\Phi^{\alpha}_{\gamma}\Phi^{\gamma}_{\beta}=0
\end{eqnarray}
Taking $<Tr\Phi>=0$ implies that $z=-\frac{3}{5}y<Tr\Phi^2>$. If we 
assume that $Diag<\Phi>=(a_1, a_2, a_3, a_4, a_5)$, then one has the
following equations:
\begin{eqnarray}
\Sigma_i a_i=0\\ \nonumber
z+2xa_i+3ya^2_i=0
\end{eqnarray}
with $i=1,...5$. Thus we have five equations and two parameters. There are
therefore three different choices for the $a_i$'s that can solve the 
above equations and they are:
\noindent Case (A):
\begin{eqnarray}
<\Phi>=0
\end{eqnarray}
In this case, SU(5) symmetry remains unbroken.
\noindent Case (B):
\begin{eqnarray}
Diag <\Phi>=(a,a,a,a,-4a)
\end{eqnarray}
In this case, SU(5) symmetry breaks down to $SU(4)\times U(1)$ and one 
can find $a=\frac{2x}{9y}$.
\noindent Case (C):
\begin{eqnarray}
Diag <\Phi>=(b, b, b,-\frac{3}{2}b, -\frac{3}{2}b)
\end{eqnarray}
This is the desired vacuum since SU(5) in this case breaks down to
$SU(3)_c\times SU(2)_L\times U(1)_Y$ gauge group of the standard model.
The value of $b=\frac{4x}{3y}$ and we choose the parameters $x$ to be 
order of $M_U$. In the supersymmetric limit all vacua are degenerate.

\bigskip

\subsection{ Low energy spectrum and doublet-triplet splitting}

\bigskip

Let us next discuss whether the MSSM arises below the GUT scale in this 
model. So far we have only obtained the gauge group. The matter content
of the MSSM is also already built into the $\bar{F}$ and $T$ multiplets.
The only remaining question is that of the two Higgs superfields $H_u$
and $H_d$ of MSSM.  They must come out of the $H$ and the $\bar{H}$
multiplets. Writing $H\equiv \left(\begin{array}{c} \zeta_u \\ H_u
\end{array} \right)$ and $\bar{H}\equiv \left(\begin{array}{c}\bar{\zeta_d}
\\ H_d\end{array}\right)$. From $W_G$ substituting the $<\Phi>$ for
case (C), we obtain, 

\begin{eqnarray}
W_{eff}= \lambda (b+M)\zeta_u\bar{\zeta_d} +\lambda(-3/2b+M) H_uH_d
\end{eqnarray}
If we choose $3/2b=M$, then the massless standard model doublets remain
and every other particle of the SU(5) model gets large mass.
The uncomfortable aspect of this procedure is that the adjustment of the 
parameters is done by hand does not emerge in a natural manner. This
procedure of splitting of the color triplets $\zeta_{u,d}$ from 
$SU(2)_L$ doublets $H_{u,d}$ is called doublet-triplet splitting and is
a generic issue in all GUT models. An advantage of SUSY GUT's is that
once the fine tuning is done at the tree level, the nonrenormalization
theorem of the SUSY models preserves this to all orders in perturbation 
theory. This is one step ahead of the corresponding situation in non-
SUSY GUT's, where the cancellation between $b$ and $M$ has to be done in 
each order of perturbation theory. A more satisfactory situation would be
where the doublet-triplet splitting emerges naturally due to requirements 
of group theory or underlying dynamics.

\bigskip
\subsection{ Fermion masses and  Proton decay}

Effective superpotential for matter sector at low energies then looks like:
\begin{eqnarray}
W_{matter} h_u QH_uu^c + h_d QH_d d^c + h_l LH_d e^c +\mu H_u H_d
\end{eqnarray}
Note that $h_d$ and $h_l$ arise from the $T\bar{F}\bar{H}$ coupling
and this satisfy the relation $h_d=h_l$. Similarly, $h_u$ arises from
the $TTH$ coupling and therefore obeys the constraint
 $h_u=h^T_u$.  (None of these constraints are present in the MSSM). The
second relation will be recognized by the reader as a partial Yukawa 
unification relation and we can therefore use the discussion of Section 2
to predict $m_{\tau}$ in terms of $m_b$. The relation between the Yuakawa 
couplings however hold for each generation and therefor imply the 
undesirable relations among the fermion masses such as $m_d/m_s=m_e/m_{\mu}$.
This relation is independent of the mass scale and therefore holds also
at the weak scale. It is in disagreement with observations by almost a 
factor of 15 or so. This a major difficulty for minimal SU(5) model. This 
problem does not reflect any fundamental difficulty with the idea of 
grand unification but rather with this particular realization. In fact
by including additional multiplets such as ${\bf 45}$ in the theory, one
can avoid this problem. Another way is to add higher dimensional 
operators to the theory such as $T\bar{F}\Phi\bar{H}/M_{Pl}$, which
can be of order of a 0.1 GeV or so and could be used to fix the muon
mass prediction from $SU(5)$.

The presence of both quarks and leptons in the same multiplet of SU(5)
model leads to proton decay. For detailed discussions of this classic
feature of GUTs, see for instance \cite{ss|mohap}. In non-SUSY SU(5),
there are two classes
of Feynman diagrams that lead to proton decay in this model: (i) the exchange
of gauge bosons familiar from non-SUSY SU(5) where effective operators
of type $e^{+\dagger}u d^{c\dagger}u$ are generated;
 and (ii) exchange of Higgs fields. In the 
supersymmetric
case theer is an additional source for proton decay coming from the
exchange of Higgsinos, where $QQH$ and $QL\bar{H}$ via $H\bar{H}$
mixing generate the effective operator $QQQL/M_H$ that leads to
proton decay. In fact, this turns out to give the dominant 
contribution.

The gauge boson exchange diagram leads to $p\rightarrow e^+\pi^0$ with
an amplitude $M_{p\rightarrow e^+\pi^0}\simeq \frac{4\pi\alpha_U}{M^2_U}$.
This leads to a prediction for the proton lifetime of:
\begin{eqnarray}
\tau_p=4.5\times 10^{29\pm.7}\left(\frac{M_U}{2.1\times 10^{14}~GeV}\right)^4
\end{eqnarray}
For $M_U\simeq 2\times 10^{16}$ GeV, one gets $\tau_p= 4.5\times 
10^{37\pm .7}$ yrs. This far beyond the capability of SuperKamiokande
experiment, whose ultimate limit is $\sim 10^{34}$ years.

Turning now to the Higgsino exchange diagram, we see that the
amplitude for this case is given by:
\begin{eqnarray}
M\simeq \frac{h_uh_d}{M_H}\cdot \frac{m_{gaugino}g^2}{16\pi^2 
M^2_{\tilde{Q}}}
\end{eqnarray}
In this formula there is only one heavy mass suppression. Although there
are other suppression factors, they are not as potent as in the gauge boson
exchange case. As result this dominates. A second aspect of this process
is that the final state is $\nu K^+$ rather than $e^+\pi^0$. This can
be seen by studying the effective operator that arises from the exchange
of the color triplet fields in the ${\bf 5}+{\bf \bar{5}}$ i.e.
$O_{\Delta B=1}= QQQL$ where $Q$ and $L$ are all superfields and are
therefore bosonic operators. In terms of the isospin and color components,
this looks like $\epsilon^{ijk}u_iu_jd_ke^-$ or $\epsilon^{ijk}u_id_jd_k\nu$.
It is then clear that unless the two $u$'s or the $d$'s in the above 
expressions belong to two different generations, the operators vanishes
due to color antisymmetry. Since the charm particles are heavier than
the protons, the only contribution comes from the second operators and the
strange quark has to be present (i.e. the operator is $\epsilon^{ijk}
u_id_js_k\nu_{\mu}$. Hence the new final state. Detailed calculations
show\cite{nath2} that for this decay lifetime to be consistent with present
observations, one must have $M_H> M_U$ by almost a factor of 10. This is 
somewhat unpleasant since it would require that some coupling in the 
superpotential has to be much larger than one. 

\bigskip

\subsection{Other aspects of SU(5)}

There are several other interesting implications of SU(5) grand unification
that makes this model attractive and testable. The model has
{\it very few parameters and hence is very predictive}. The MSSM has got
more than a hundred free parameters, that makes such models expertimentally
quite fearsome and of course hard to test. On the other hand, once the model
is embedded into SUSY SU(5) with Polonyi type supergravity, the number of 
parameters reduces to just five: they are the $A,~ B, ~m_{3/2}$ which
parameterize the effects of supergravity discussed in section I, $~\mu$
parameter which is the $H_uH_d$ mixing term in the superpotential 
also present in the superpotential and $m_{\lambda}$, the universal
gaugino mass. This reduction in the number of parameters has the following
implications:

\bigskip

\noindent{\it (i) Gaugino unification}:

\bigskip

At the GUT scale, we have the three gaugino masses equal (i.e.
$m_{\lambda_1}=m_{\lambda_2}=m_{\lambda_3}$. There value at the weak 
scale can be predicted by using the RG running as follows:
\begin{eqnarray}
\frac{dm_{\lambda_i}}{dt}=\frac{b_i}{2\pi}\alpha_im_{\lambda_i}
\end{eqnarray} 
Solving these equations , one finds that at the weak scale, we have
\begin{eqnarray}
m_{\lambda_1} : m_{\lambda_2} : m_{\lambda_3} = \alpha_1 :\alpha_2 :\alpha_3
\end{eqnarray}
Thus discovery of gaugino's will test this formula and therefore SU(5)
grand unification.

\bigskip

\noindent{\it (ii) Prediction for squark and slepton masses}

\bigskip

At the supersymmetry breaking scale, all scalar masses in the simple
supergravity schemes are equal. Again, one can predict their weak scale
values by the RGE extrapolation. One finds the following 
formulae\cite{nath3}: 
\begin{eqnarray}
m^2_{\tilde{Q}}=m^2_{3/2}+m^2_{Q}+\frac{\alpha_U}{4\pi}[\frac{8}{3}f_3
+\frac{3}{2}f_2+\frac{1}{30}f_1]m^2_{\lambda_U}+Q^Z_{Q}M^2_Zcos^22\beta
\end{eqnarray}
where $Q^Z_u=\frac{1}{2}-\frac{2}{3}sin^2\theta_W$ and
$Q^Z_d=-\frac{1}{2}+\frac{1}{3}sin^2\theta_W$ and
$f_k=\frac{t(2-b_kt)}{1+b_kt^2}$ and $b_k$ are the coefficients of the
RGE's for coupling constant evolutions given earlier. A very obvious 
formula for the sleptons can be written down. It omits the strong coupling
factor. A rough estimate gives that $m^2_{\tilde{l}}\simeq m^2_{3/2}$
and $m^2_{\tilde{Q}}\simeq m^2_{3/2}+ 4 m^2_{\lambda_U}$. This could
therefore serve as independent tests of the SUSY SU(5).

\bigskip

\subsection{Problems and prospects for SUSY SU(5)}

While the simple SUSY SU(5) model exemplifies the power and utility of
the idea of SUSY GUTs, it also brings to the surface some of the problems
one must solve if the idea eventually has to be useful. Let us enumerate
them one by one and also discuss the various ideas proposed to overcome
them.

\bigskip

\noindent {\underline{{\it (i) R-parity breaking:} }

\bigskip

There are renormalizable terms in the superpotential that break baryon
and lepton number:
\begin{eqnarray}
W'=\lambda_{abc} T_a\bar{F}_b\bar{F}_c
\end{eqnarray}
When written in terms of the component fields, this leads to R-parity
breaking terms of the MSSM such as $L_aL_be^c_c$, $QLd^c$ as well as
$u^cd^cd^c$ etc. The new point that results from grand unification is
that there is only one coupling parameter that describes all three
types of terms and also the coupling $\lambda$ satisfies the antisymmetry
in the two generation indices $b,c$. This total number of parameters
that break R-parity are nine instead of 45 in the MSSM.
There are also nonrenormalizable terms of the form 
$T\bar{F}\bar{F}(\Phi/M_{P\ell})^n$\cite{zura}, which are significant
for $n=1,2,3,4$ and can add different complexion to the R-parity violation.
 Thus, the SUSY
SU(5) model does not lead to an LSP that is naturally stable to lead to
a CDM candidate. As we will see in the next section, the SO(10) model
provides a natural solution to this problem if only certain Higgs superfields
are chosen.
\bigskip

\noindent{\underline{\it (ii) Doublet-triplet splitting problem:}}

\bigskip

We saw earlier that to generate the light doublets of the MSSM, one needs
a fine tuning between the two parameters $3/2\lambda b$
 and $M$ in the superpotential.
However once SUSY breaking is implemented via the 
hidden sector mechanism
one gets a SUSY breaking Lagrangian of the form:
\begin{eqnarray}
L_{SB}=A\lambda\bar{H}\Phi H +BM \bar{H}H+ h.c.
\end{eqnarray}
where the symbols in this equation are only the 
scalar components of the
superfields. In general supergravity scenarios, 
$A\neq B$. As a result,
when the Higgsinos are fine tuned to have mass in 
the weak scale range,
the same fine tuning does not leave the scalar 
doublets at the weak scale.

There are two possible  ways out of this problem: 
we discuss them below.

\bigskip

\noindent{\it (iiA) Sliding singlet}
\bigskip

The first way out of this is to introduce a 
singlet field $S$ and choose the
superpotential of the form:
\begin{eqnarray}
W_{DT}=2 \bar{H}\Phi H+ S \bar{H} H
\end{eqnarray}
The supersymmetric minimum of this theory is given 
by:
\begin{eqnarray}
F_H=H_u(-3b+<S>)=0
\end{eqnarray}
The $F_{\zeta}$ equation is automatically 
satisfied when color is
unbroken as is required to make the theory 
physically acceptable.
We then see that one then automatically gets 
$<S>=3 b$
which is precisely the condition that keeps the 
doublets light.
Thus the doublets remain naturally of the weak 
scale without any need for
fine tuning. This is called the sliding singlet 
mechanism. In this case
the supersymmetry breaking at the tree level maintains the massless of   
the MSSM doublets for both the fermion as well as the bosonic components.
There is however a problem that arises once one loop corrections are
included- because they lead to corrections for the $<S>$ vev of order
$\frac{1}{16\pi^2}m_{3/2}M_U$ which then produces a mismatch in the
cancellation of the bosonic Higgs masses. One is back to square one!

\bigskip
\noindent{\it (iiB) Missing partner mechanism:}
\bigskip

A second mechanism that works better than the previous one is the so called
missing partner mechanism where one chooses to break the GUT symmetry by
a multiplet that has coupling to the $H$ and $\bar{H}$ and other multiplets
in such a way that once SU(5) symmetry is broken, only the color triplets
in them have multiplets in the field it couples to pair up with but not
weak doublets. As a result, the doiublet naturally light. An example is
provided by adding the {\bf 50}, $\bar{\bf 50}$ (denoted by 
$\Theta^{\alpha\beta}_{\gamma\delta\sigma}$ and $\bar{\Theta}$ respectively)
and replacing {\bf 24} by the {\bf 75} (denoted $\Sigma$) dimensional 
multiplet.
Note that {\bf 75} dim multiplet has a standard model singlet in it so
that it breaks the SU(5) down to the standard model gauge group. At the same
time {\bf 50} has a color triplet only and no doublet. The ${\bf 50 .
75.\bar{5}}$
coupling enables the color triplet in {\bf 50} and $\bar{\bf 5}$ to pair up
leaving the weak doublet in $\bar{H}$ light. The superpotential in this 
case can be given by
\begin{eqnarray}
W_G= \lambda_1 \Theta \Sigma H + \lambda_2 \bar{\Theta} \Sigma \bar{H}
+ M\Theta\bar{\Theta} +f(\Sigma)
\end{eqnarray}

 This mechanism can be applied
in the case of other groups too.

\bigskip
\noindent{\underline{\it (iii) Baryogenesis problem}}

\bigskip

There are also other problems with the SUSY SU(5) model that suggest
that other
GUT groups be considered. One of them is the problem with generating
the baryon asymmetry of the universe in a simple manner. The point is that
if baryon asymmetry in this model is generated at the GUT scale as 
is customarily done, then the there must also simultanoeusly be a 
lepton asymmetry such that $B-L$ symmetry is preserved. The reason for 
this is that
all interactions of the simple SUSY models conserve B-L symmetry.
As a result, we can write the 
$n_B=\frac{1}{2}n_{B-L}+\frac{1}{2}n_{B+L}
=\frac{1}{2}n_{B+L}$.  The
problem then is that the sphaleron interactions\cite{kuzmin}
which are in equilibrium for $ 10^{2}~GeV\leq T \leq 10^{12}~GeV$,
will erase the $n_{B+L}$ since they violate the $B+L$ quantum 
number.
Thus the GUT scale baryon asymmetry cannot survive below the weak 
scale.   
Of course one could perhaps generate baryons at the weak scale 
using the
sphaleron processes. But no simple and convincing mechanism seems to 
have been in place yet. Thus it may be wise to look at higher 
unification groups.

\bigskip
\noindent{\underline{\it (iv) Neutrino masses}}

\bigskip

Finally, in the SU(5) model there seems to no natural mechanism for
generating neutrino masses although using the R-parity violating
interactions for such a purpose has often been suggested. One would
then have to accept the required smallness of their couplings has to
be put in by hand.

\bigskip
\noindent{\underline{\it (v) Vacuum degeneracy and supergravity effects}}
\bigskip

A generic cosmological problem of most SUSY GUT's is the vacuum degeneracy
obtained in the case of the SU(5) model in the supersymmetric limit discussed
in section 3.2 above. Recall that SU(5) symmetry breaking via the {\bf 24} 
Higgs superfield leaves three vacua i.e. the SU(5) , $SU(4)\times U(1)$ 
and the $SU(3)_c\times SU(2)_L\times U(1)_Y$ ones with same vacuum energy.
The question then is how does the universe settle down to the standard model
vacuum. It turns out that once the supergravity effects are included, the
three vacua have different energies coming from the 
$\frac{-3}{M^2_{Pl}}|W|^2$ term in the effective bosonic potential. Using the
values of the parameters a and b above that characterise the vacua, we
find these energies to be:
\begin{eqnarray}
<\Phi>=0:  V_0=0\\ \nonumber
SU(4)\times U(1): 
V_0=-3\left(\frac{80}{243}\right)^2\frac{x^6}{M^2_{Pl}y^4}\\ \nonumber
SU(3)_c\times SU(2)_L\times U(1)_y: 
V_0=-3\frac{1600}{81}\frac{x^6}{M^2_{Pl}y^4}
\end{eqnarray}
This would appear quite intereseting since indeed the standard model 
vacuum has the lowest vacuum energy. However that is misleading since this
evaluation is done prior to the setting of the cosmological constant to zero.
Once that is done, the standard model indeed acquires the highest vacuum 
energy. Thus this remains a problem. One way to avoid this would be to 
imagine that the standard model is indeed stuck in the wrong vacuum but
the tunneling probability to other vacua is negligible or at least it is 
such that the tunnelling time is longer than the age of the universe.

It is worth pointing out that in the case where the SU(5) symmetry is 
broken by the {\bf 75} dim. multiplet, theer is no $SU(4)\times U(1)$
inv. vacuum. Similarly one can imagine eliminating the SU(5) inv vacuum 
by adding to the superpotential terms like $S(\Sigma^2-M^2_U)$.

\section{Supersymmetric SO(10)}

In this section, we like to discuss supersymmetric SO(10) models which
have a number of additional desirable features over SU(5) model. For
instance, all the matter fermions fit into one spinor representation of 
SO(10); secondly, the SO(10) spinor being 16-dimensional, it contains the 
right-handed neutrino leading to nonzero neutrino masses. The gauge group
of SO(10) is left-right symmetric which has the consequence that it can 
solve the SUSY CP problem and R-parity problem etc. the MSSM unlike the SU(5)
model. Before proceeding to a discussion of the model, let us briefly
discuss the group theory of SO(10).

\subsection{Group theory of SO(10)}

 The SO(2N) group is defined by the Clifford algebra of 2N elements,
$\Gamma_a$ which satisfy the following anti-commutation relations:
\begin{eqnarray}
[\Gamma_a,\Gamma_b]_+=2 \delta_{ab}
\end{eqnarray}
where $a,b$ go from 1...2N.
The generators of SO(2N) group are then given by $\Sigma_{ab}\equiv 
\frac{1}{2i} [\Gamma_a,\Gamma_b]_-$. The study of the spinor representations
and simple group theoretical manipulations with SO(2N) is considerably
simplified if one uses the SU(N) basis for SO(2N)\cite{sakita}.   

To discuss the SU(N) basis, let us introduce N anticommuting operators
$\chi_i$ and $\chi^{\dagger}_i$ satisfying the following anticommuting
relations:
\begin{eqnarray}
[\chi_i,\chi^{\dagger}_j]_+=\delta_{ij}
\end{eqnarray}
We can then express the elements of the Clifford algebra $\Gamma_a$'s
in terms of these fermionic operators as follows:
\begin{eqnarray}
\Gamma_{2i-1}=\frac{\chi_i-\chi^{\dagger}_i}{2i}\\ \nonumber
\Gamma_{2i}=\frac{\chi_i+\chi^{\dagger}_i}{2}
\end{eqnarray}

The spinor representations of the SO(10) group can be obtained
using this formalism as follows:
\begin{eqnarray}
\Psi=\left(\begin{array}{c}
\chi^{\dagger}_j|0> \\ \chi^{\dagger}_j\chi^{\dagger}_k\chi^{\dagger}_l|0>\\
\chi^{\dagger}_j\chi^{\dagger}_i\chi^{\dagger}_l\chi^{\dagger}_m
\chi^{\dagger}_n|0>\end{array} \right)
\end{eqnarray}
 By simple counting, one can see that this is a {\bf 16} dimensional
representation. The states in the {\bf 16}-dim. spinor have the right 
quantum numbers to accomodate the matter fermins of one generation. The 
different particle states can be easily 
identified: e.g. $e^-=\chi^{\dagger}_4|0>$; $ d^c_i=\chi^{\dagger}_i|0>; 
u_i=\chi^{\dagger}_2\chi^{\dagger}_3\chi^{\dagger}_5|0>; 
e^+=\chi^{\dagger}_1\chi^{\dagger}_2\chi^{\dagger}_3|0>$ etc.

 Other representations such as {\bf 10} are given simply
by the $\Gamma_a$, {\bf 45} by $[\Gamma_a, \Gamma_b]$ etc.
In other words, they can be denoted by vectors with totally antisymmetric
indices: The tensor representations that will be necessary in our discussion
are ${\bf 10}\equiv H_a$; ${\bf 45}\equiv A_{ab}$, ${\bf 120}\equiv 
\Lambda_{abc}$, ${\bf 210}\equiv \Sigma_{abcd}$ and ${\bf 126}\equiv 
\Delta_{abcde}$. (All indices here are totally antisymmetric).
One needs a charge conjugation operator to write Yukawa couplings
such as $\Psi\Psi H$ where $H\equiv {\bf 10}$.  It is given by   
$C\equiv\Pi_i \Gamma_{2i-1}$ with $i=1,...5$.
The generators of SU(4) and $SU(2)_L\times SU(2)_R$ can be written down
in terms of the $\chi$'s. The fact that SU(4) is isomorphic to  
SO(6) implies that the generators of SU(4) will involve only $\chi_i$ and 
its    
hermitean conjugate for $i=1,2,3$ whereas the $SU(2)_L\times SU(2)_R$ 
involves
only $\chi_p$  (and its h.c.) for $p=4,5$. The $SU(2)_L$ generators are:
$I^+_L=\chi^{\dagger}_4\chi_5$ and $I^-_L$ and $I_{3,L}$ can be found 
from it.
Similarly, $I^+_R=\chi^{\dagger}_5\chi^{\dagger}_4$ and the other right
handed generators can be found from it.
 For instance $I_{3R}=\frac{1}{2}[I_{+R}-I_{-R}]$ etc. We also have
\begin{eqnarray}
B-L=-\frac{1}{3}\Sigma_i\chi^{\dagger}_i\chi_i+
\Sigma_p\chi^{\dagger}_p\chi_p
\\ \nonumber
Q=\frac{1}{3}\Sigma_i\chi^{\dagger}_i\chi_i -\chi^{\dagger}_4\chi_4
\end{eqnarray}
 
This formulation is one of many ways one can deal with the group theory
of SO(2N)\cite{zee}. An advantage of of the spinor basis is that 
calculations such as those for ${\bf 16}.{\bf 10}. {\bf 16}$  
need only manipulations of the anticommutation relations among the     
$\chi_i$'s and bypass any matrix multiplication.

As an example, suppose we want to evaluate up and down quark masses
induced by the weak scale vev's from the {\bf 10} higgs. We have to
evaluate $\Psi C \Gamma_a \Psi H_a$. To see which components of H  
corresponds to electroweak doublets, let us note that $SO(10)\rightarrow
SO(6)\times SO(4)$; denote $a=1,..6$ as the SO(6) indices and $p=7..10$ 
as the SO(4) indices. Now SO(6) is isomorphic to SU(4) which we identify as
SU(4) color with lepton number as fourth color\cite{pati} and SO(4) is
isomorphic to $SU(2)_L\times SU(2)_R$ group. To evaluate the above
matrix element, we need to give vev to $H_{9,10}$ since all other elements
have electric charge. This can be seen from the SU(5) basis, where
$\chi_5$ , corresponding to the neutrino has zero charge whereas all
the other $\chi$'s have electric charge as can be seen from the
formula for electric charge in terms of $\chi$'s given above. Thus all 
one needs to evaluate is typically a matrix element of the type
$<0|\chi_1\Gamma_9 C \chi^{\dagger}_2\chi^{\dagger}_3\chi^{\dagger}_4|0>$.
In this matrix element, only terms $\chi_5$ from $\Gamma_9$ and
$\chi_2\chi_3\chi_4\chi^{\dagger}_1\chi^{\dagger}_5$ will contribute 
and yield a value one.

\bigskip
\subsection{Symmetry breaking and fermion masses}
\bigskip

Let us now proceed to discuss the breaking of SO(10) down to the standard
model. SO(10) contains the maximal subgroups $SU(5)\times U(1)$ and
$SU(4)_c\times SU(2)_L\times SU(2)_R\times Z_2$ where the $Z_2$ group
corresponds to charge conjugation. The $SU(4)_c$ group contains the
subgroup $SU(3)_c\times U(1)_{B-L}$. Before discussing the symmetry 
breaking, let us digress to discuss the $Z_2$ subgroup and its implications. 

The discrete subgroup $Z_2$ is often called
D-parity in literature\cite{parida}. Under D-parity, $u\rightarrow u^c; 
e\rightarrow e^c$ etc. In general the D-parity symmetry and 
the $SU(2)_R$ symmetry can be broken separately from each other. This has 
several interesting physical implications. For example if D-parity breaks at
a scale ($M_P$) higher than $SU(2)_R$ ($M_R$) (i.e. $M_P>M_R$), then the
Higgs boson spectrum gets asymmetrized and as a result, the two gauge 
couplings evolve in a different manner. At $M_R$, one has $g_L\neq g_R$.
The SO(10) operator that implements the D-parity operation is given by
$D\equiv \Gamma_2\Gamma_3\Gamma_6\Gamma_7$. The presence of D-parity group
below the GUT scale can lead to formation of domain walls bounded by strings
\cite{shafi}. This can be cosmological disaster if $M_P=M_R$\cite{shafi}
whereas this problem can be avoided if\cite{parida} $M_P > M_R$. Another
way to avoid such problem will be to invoke inflation with a reheating 
temperature $T_R\leq M_R$. 

There are therefore many ways to
break SO(10) down to the standard model. Below we list a few of the
interesting breaking chains along with the SO(10) multiplets whose vev's lead
to that pattern.

\bigskip
\noindent (A) $SO(10)\rightarrow SU(5)\rightarrow G_{STD}$
\bigskip

The Higgs multiplet responsible for the breaking at the first stage is
a {\bf 16} dimensional multiplet (to be denoted $\psi_H$) which has a
field with the quantum number of $\nu^c$ which is an SU(5) singlet but 
with non-zero $B-L$ quantum number. The second stage can be achieved by 
\begin{eqnarray}
{\bf 16_H}\rightarrow {\bf 1}_{-5}+{\bf 10}_{-1}+{\bf {\bar 5}}_{+3}
\end{eqnarray}
The breaking of the SU(5) group down to the standard model is implemented
by the {\bf 45}-dimensional multiplet which contains the {\bf 24} dim.
representation of SU(5), which as we saw in the previous section
contains a singlet of the standard model group. In the matrix notation,
we can write breaking by {\bf 45} as $<A>=i\tau_2\times Diag(a,a,a,b,b,)$
where $a\neq 0$ whereas we could have $b=0$ or nonzero.

A second symmetry breaking chain of physical interest is:

\bigskip
\noindent (B) $ SO(10)\rightarrow G_{224D}\rightarrow G_{STD}$
\bigskip

\noindent where we have denoted $G_{224D}\equiv SU(2)_L\times SU(2)_R\times
SU(4)_c\times Z_2$. We will use this obvoius shorthand for the different
 subgroups. This breakingis achieved by the Higgs multiplet
\begin{eqnarray}
{\bf 54}= (1,1,1)+ (3,3,1)+(1,1,20')+(2,2,6)
\end{eqnarray}
The second stage of the breaking of $G_{224D}$ down to $G_{STD}$ is achieved
in one of two ways and the physics in both cases are very different as we
will see later: (i) {\bf 16}+${\bf \bar{16}}$ or (ii) {\bf 126}+
${\bf {\bar 126}}$. For clarity, let us give the $G_{224D}$ decomposition of
the {\bf 16} and {\bf 126}.
\begin{eqnarray}
{\bf 16}= (2,1,4)+(1,2,\bar {4})\\ \nonumber
{\bf 126}= (3,1,10)+(1,3,\bar{10})+(2,2,15)+(1,1,6)
\end{eqnarray}
In matrix notation, we have 
\begin{eqnarray}
{<\bf 54>}=Diag (2a,2a,2a,2a,2a,2a,-3a,-3a,-3a,-3a)
\end{eqnarray}
and for the {\bf 126} case it is the
$\nu^c\nu^c$ component that has nonzero vev.

It is important to point out that since the supersymmetry has to be
maintained down to the electroweak scale, we must consider the Higgs
bosons that reduce the rank of the group in pairs (such as {\bf 16}+
${\bf {\bar 16}}$). Then the D-terms will cancel among themselves. However,
such a requirement does not apply
if a particular Higgs boson vev does not reduce the rank.

\bigskip
\noindent (C) $ SO(10)\rightarrow G_{2231}\rightarrow G_{STD} $
\bigskip

\noindent This breaking is achieved by a combination of {\bf 54} and {\bf 45}
dimensional Higgs representations. Note the absence of the $Z_2$ symmetry 
after the first stage of breaking. This is because the (1,1,15) (under 
$G_{224}$) submultiplet that breaks the SO(10) symmetry is odd under the
 D-parity. The second stage breaking is as in the case (B).

\bigskip
\noindent (D) $SO(10)\rightarrow G_{224}\rightarrow G_{STD}$
\bigskip

\noindent Note the absence of the D-parity in the second stage. This is
achieved by the Higgs multiplet {\bf 210} which decomposes under $G_{224}$
as follows:
\begin{eqnarray}
{\bf 210}= (1,1,15)+(1,1,1)+(2,2,10)\\ \nonumber
+(2,2,\bar 10)+(1,3,15)+(3,1,15) +(2,2,6)
\end{eqnarray}
The component that acquires vev is $<\Sigma_{78910}>\neq 0$.

It is important to point out that since the supersymmetry has to be 
maintained down to the electroweak scale, we must consider the Higgs
bosons that reduce the rank of the group in pairs (such as {\bf 16}+
${\bf {\bar 16}}$). Then the D-terms will cancel among themselves. However,
such a requirement does not apply
if a particular Higgs boson vev does not reduce the rank.

Let us now proceed to the discussion of fermion masses. As in all gauge 
models, they will arise out of the Yukawa couplings after spontaneous 
symmetry breaking. To obtain the Yukawa couplings, we first note that

\noindent ${\bf 16\times 16=10 +120 +126}$.
 Therefore the gauge invariant couplings are
of the form ${\bf 16. 16 .10}\equiv \Psi^TC^{-1}\Gamma_a\Psi H_a$;
${\bf 16. 16 .120}\equiv \Psi \Gamma_a\Gamma_b\Gamma_c\Psi \Lambda_{abc}$
and ${\bf 16.16.\bar{126}}\equiv \Psi 
\Gamma_a\Gamma_b\Gamma_c\Gamma_d\Gamma_e\Psi \bar{\Delta}_{abcde}$.
We have suppressed the generation indices. Treating the Yukawa couplings as
matrices in the generation space, one gets the following symmetry 
properties for them: $h_{10}=h^T_{10}$; $h_{120}=-h^T_{120}$ and 
$h_{126}=h^T_{126}$ where the subscripts denote the Yukawa couplings
of the spinors with the respective Higgs fields.

To obtain fermion masses after electroweak symmetry breaking, one has to 
give vevs to the following components of the fields in different cases:
$<H_{9,10}>\neq 0$; $\Lambda_{789,7810}\neq 0$ or $\Lambda_{129}
=\Lambda_{349}=\Lambda_{569}\neq 0$ (or with 9 replaced by 10) and similarly
$\Delta_{12789}=\Delta_{34789}=\Delta_{56789}\neq 0$ etc. Several important
constraints on fermion masses implied in the SO(10) model are:

\noindent (i) If there is only one {\bf 10} Higgs responsible for the
masses, then only $<H_{10}>\neq 0$ ane has the relation 
$M_u=M_d=M_e=M_{\nu^D}$; where the $M_F$ denote the mass matrix for 
the F-type fermion.

\noindent (ii) If there are two {\bf 10}'s, then one has $M_d=M_e$ and 
$M_u=M_{\nu^D}$.

\noindent(iii) If the fermion masses are generated by a {\bf 126}, then
we have the mass relation following from $SU(4)$ symmetry i.e.
$3M_d=-M_e$ and $3M_u=-M_{\nu^D}$.

It is then clear that, if we have only {\bf 10}'s generating fermion masses
we have the bad
mass relations for the first wo generations in the down-electron sector. On
the other hand it provides the good $b-\tau$ relation. One way to cure
it would be to bring in contributions from the {\bf 126}, which split
the quark masses from the lepton masses- since in the $G_{224}$ language,
it contains $(2,2,15)$ component which gives the mass relation $m_e=
-3 m_d$. This combined with the {\bf 10} contribution can perhaps
provide phenomenologically viable fermion masses. With this in mind,
we note the suggestion of Georgi and Jarlskog\cite{jarl}
who proposed that
one should have the $M_d$ and $M_e$ of the following forms to avoid the
bad mass relations among the first generations while keeping $b-\tau$           
unification:
\begin{eqnarray}
M_d=\left(\begin{array}{ccc}
0 & d & 0\\
d & f & 0\\
0 & 0 & g\end{array}\right);
M_e=\left(\begin{array}{ccc}
0 & d & 0\\
d & -3f & 0\\
0 & 0 & g\end{array} \right)
\end{eqnarray}
\begin{eqnarray}
M_u=\left(\begin{array}{ccc}
0 & a & 0\\
a & 0 & b\\
0 & b & c \end{array}\right)
\end{eqnarray}
These mass matrices lead to $m_b=m_{\tau}$ at the GUT scale and
$\frac{m_e}{m_{\mu}}\simeq \frac{1}{9}\frac{m_d}{m_s}$ which are
in much better agreement with observations. There have been many
derivations and analyses of 
these mass matrices in the context of SO(10) models\cite{he}

\bigskip
\subsection{Neutrino masses, R-parity breaking, {\bf 126} vrs. {\bf 16}:} 
\bigskip

One of the attractive aspects of the SO(10) models is the left-right
symmetry inherent in the model. A consequence of this is the complete 
quark-lepton symmetry in the spectrum. This implies the existence of the
right-handed neutrino which as we will see is crucial to our understanding 
of the small neutrino masses. This comes about via the see-saw mechanism 
mentioned earlier in section 1. The generic see-saw mechanism for one 
generation can be seen in the context of the standard model with the
inclusion of an extra righthanded neutrino which is a singlet of the
standard model group. As is easy to see, if there is a right-handed
neutrino denoted as $\nu^c$, then we have additional terms in the 
MSSM superpotential of the form $h_-{\nu}LH_u\nu^c+ M\nu^c\nu^c$. After
electroeak symmetry breaking, there emerges a $2\times 2$ mass matrix for
the $(\nu, \nu^c)$ system of the following form:
\begin{eqnarray}
M_{\nu}=\left(\begin{array}{cc}
0 & h_uv_u\\
h^T_uv_u & M\end{array}\right)
\end{eqnarray}
This matrix can be diagonalized easily and noting that $M\gg h_uv_u$, we 
find a light eigenvalue $m_{\nu}\simeq \frac{(h_uv_u)^2}{M}$ and a heavy 
eigenvalue $\simeq M$. The light eigenstate is predominantly the light
weakly interacting neutrino and the heavy eigenstate is the superweakly
interacting right handed neutrino. Thus without any fine tuning, one sees
(using the fact that $m_f\simeq h_uv_u$) that $m_{\nu}\simeq m^2_f/ M\ll 
m_f$.
This is known as the see-saw mechanism\cite{ramond}. For future reference,
we note that $h_uv_u$ for three generations is a matrix and is called
the Dirac mass of the neutrrino. The left as well as the right handed
neutrinos in this case are Majorana neutrinos i.e. they are self
conjugate. For detailed discussion the Majorana masses, see Ref.\cite{pal}.

While in the context of the standard model it is natural to expect 
$M\gg v_u$, we cannot tell what the value of M is; secondly, the approximation
of $h_uv_u\simeq m_f$ is also a guess. The SO(10) model has the potential
to make more quantitative statements about both these aspects of the see-saw
mechanism.

To see the implications of embedding see-saw matrix in the SO(10) model, 
let us first note that
if the only source for the quark and charged lepton masses is the {\bf 10}-
dim. rep. of SO(10), then we have a relation between the Dirac mass of the
neutrino and the up quark masses: $M_u=M_{\nu^D}$. Let us now note that
the $\nu^c\nu^c$ mass term $M$ arises from the vacuum expectation value (vev)
of the $\nu^c\nu^c$ component of $ {\bf \bar 126}$ and therefore corresponds
to a fundamental gauge symmetry breaking scale in the theory which can be
determined from the unification hypothesis. Thus apart from the coupling 
matrix of the ${\bf \bar{126}}$ denoted by $f$, everything can be determined.
This gives predictive power to the SO(10) model in the neutrino sector. For
instance, if we take typical values for the $f$ coupling to be one and
ignore the mixing among generations, then, we get
\begin{eqnarray}
m_{\nu_e}\simeq m^2_u/10 fv_{B-L}\\ \nonumber
m_{\nu_{\mu}}\simeq m^2_c/10 fv_{B-L}\\ \nonumber
m_{\nu_{\tau}}\simeq m^2_t/10 fv_{B-L}
\end{eqnarray}
If we take $v_{B-L}\simeq 10^{12}$ GeV, then we get, $m_{\nu_e}\simeq
10^{-8}$ eV; $m_{\nu_{\mu}}\simeq 10^{-4}$ eV and $m_{\nu_{\tau}}\simeq
$ eV. These values for the neutrino masses are of great deal of interest
in connection with the solutions to the solar neutrino problem as well 
as to the hot dark matter of the universe. Things in the SO(10) model
are therefore so attractive that one can go further in this discussion and
about the prediction for $v_{B-L}$ in the SO(10) model. The situation here
however is more complex and we summarize the situation below.

If the particle spectrum all the way until the GUT scale is that of the
MSSM, then both the $M_U$ and the $v_{B-L}$ are same and $\simeq 2\times
10^{16}$ GeV. On the other hand, if above the $v_{B-L}$ scale, the symmetry
is $G_{2213}$ and the spectrum has two bidoublets of the SUSYLR theory,
$B-L=\pm 2$ triplets of both the left and the right handed groups and a
color octet, then one can easily see that in the one loop approximation,
the $v_{B-L}\simeq 10^{13}$ GeV or so. On the other hand with a slightly
more complex system described in section 2, we could get $v_{B-L}$ almost
down to a few TeV's.
 Thus unfortunately, the magnitude of the scale $v_{B-L}$ is quite model
dependent. 

Finally, it is also worth pointing out that the equality of $M_u$ and 
$M_{\nu^D}$ is not true in more realistic models. The reason is that
if the charged fermion masses arise purely from the {\bf 10}-dim.
rep.s than one has the undesirable relation $m_d/m_s=m_e/m_{\mu}$ which
was recognized in the SU(5) model to be in contradiction with observations.
Therefore in order to predict neutrino masses in the SO(10) model, one needs
additional assumptions than simply the hypothesis of grand unification.

\bigskip
\noindent{\it (i) Neutrino masses in the case $B-L$ breaking by ${\bf 
16}_H$:} \bigskip

As has been noted, it is possible to break $B-L$ symmetry in the
SO(10) model by using the ${\bf 16}+\bar{\bf 16}$ pair. This line of 
model building has been inspired by string models which in the old
fashioned fermionic compactification do not seem to lead to {\bf 126}
type representations\cite{dienes} at any level\cite{lykken}.
There have been several realistic models constructed along these
lines\cite{hall}. In this case, one must use higher dimensional
operators to get the $\nu^c$ mass. For instance the operator
${\bf 16_m 16_m \bar{16}_H\bar{16}_H}/M_{Pl}$ after $B-L$ breaking
would give rise to a $\nu^c$ mass $\sim v^2_{B-L}/M_{Pl}$. For $v_{B-L}
\simeq M_U$, this will lead to $M_{\nu^c}\simeq 10^{13}$ GeV. This then 
leads to the neutrino spectrum of the above type.

Another way to get small neutrino masses in SO(10) models with {\bf 16}'s
rather than {\bf 126}'s without invoking higher dimensional operators is
to use the 3$\times $3 see-saw\cite{valle} rather than the two by two one
discussed above. To implement the 3X3 see-saw, one needs an extra singlet
fermion and write the following superpotential:
\begin{eqnarray}
W_{33}= h\Psi H \Psi+f\Psi \bar{\Psi}_HS +\mu S^2
\end{eqnarray}
After symmetry breaking, one gets the following mass matrix in the basis
$(\nu,\nu^c,S)$:
\begin{eqnarray}
M_{\nu}=\left(\begin{array}{ccc}
0 & hv_u & 0\\
hv_u & 0 & f\bar{v}_R\\
0 & f\bar{v}_R & \mu \end{array}\right)
\end{eqnarray}
where $\bar{v}_R$ is the vev of the $\bar {\bf 16}_H$. On diagonalizing
for the case $v_u\simeq \mu\ll v_R$, one finds the lightest neutrino
mass to be $m_{\nu}\simeq \frac{\mu h^2v^2_u}{fv_R}$ and two other
heavy eigenstates with masses of order $fv_R$.

\bigskip
\noindent {\it (ii) R-parity conservation: automatic vrs. enforced:}
\bigskip

One distinct advantage of {\bf 126} over {\bf 16} is in the property that
the former leads to a theory that conserves R-parity automatically
even after $B-L$ symmetry is broken. This is very easy to see as was
emphasized in section 1. Recall that $R=(-1)^{3(B-L)+2S}$. Since the
{\bf 126} breaks $B-L$ symmetry via the $\nu^c\nu^c$ component, it obeys
the selection rule $B-L=2$. Putting this in the formula for $R$, we see
clearly that R-parity remains exact even after symmetry breaking. On the
 other hand, when {\bf 16} is employed, $B-L$ is broken by the $\nu^c$
component which has $B-L=1$. As a result R-parity is broken after symmetry
breaking. To see some explicit examples, note that with ${\bf 16}_H$, one
can write renormalizable operators in the superpotential of the form
$\Psi\Psi_H H$ which after $<\nu^c>\neq 0$ leads to R-parity breaking terms of
the form $LH_u$ discussed in the sec.1. When  one goes to the nonrenormalizable
operators many other examples arise: e.g. $ \Psi\Psi\Psi\Psi_H/M_{Pl}$
after symmetry breaking lead to $QLd^c,LLe^c$ as well as $u^cd^cd^c$
type terms.

\bigskip
\subsection{Doublet-triplet splitting (D-T-S):}

As we noted in sec.3, splitting the weak doublets from the color
triplets appearing in the same multiplet of the GUT group is a very
generic problem of all grand unification theories. Since in the SO(10)
 models, the fermion masses are sensitive to the GUT multiplets which
lead to the low energy doublets, the problem of D-T-S acquires an added
complexity. What we mean is the following: as noted earlier, if there
are only {\bf 10} Higgses giving fermion masses, then we have the bad
relation $m_e/m_{\mu}=m_d/m_s$ that contradicts observations. One way to cure
this is to have either an {\bf 126} which leaves a doublet from it in the
low energy MSSM in conjunction with the doublet from the {\bf 10}'s
or to have only {\bf 10}'s and have non-renormalizable operators
give an effective operator which transforms like {\bf 126}. This means
that the process of doublet triplet splitting must be done in a way
that accomplishes this goal.

One of the simplest ways to implement D-T-S is to employ the missing vev
mechanis\cite{dimo}, where one takes two {\bf 10}'s (denoted by $H_{1,2}$)
and couple them to the {\bf 45} as $AH_1H_2$. If one then gives vev to
$A$ as $<A>=i\tau_2\times Diag(a,a,a,0,0)$, then it is easy to varify that
the doublets (four of them) remain light. This model without further
ado does not lead to MSSM. So one must somehow make two of the four doublets
heavy. This was discussed in great detail by Babu and Barr\cite{dimo}.
A second problem also tackled by Babu and Barr is the question that
once the SO(10) model is made realistic by the addition of say $\bf 16+
\bar{16}$ , then new couplings of the form $\bf 16. \bar{16}.  45$ exist in the
theory that give nonzero entries at the missing vev position thus
destroying the whole suggestion. There are however solutions to this
problem by increasing the number of {\bf 45}'s.

Another more practical problem with this method is the following.
As mentioned before, the low energy doublets in this method are coming from 
{\bf 10}'s only and is problematic for fermion mass relations. This
problem was tackled in two papers\cite{lee,babu2}. In the first paper,
it was shown how one can mix in a doublet from the {\bf 126} so that
the bad fermion mass relation can be corrected. To show the bare essentials
of this techniques, let consider a model with a single $H$, single pair
$\Delta+\bar{\Delta}$ and a $A$ and $S\equiv {\bf 54}$ and write the
 following superpotential:
\begin{eqnarray}
W= M\Delta\bar{\Delta} +\Delta A\bar{\Delta} +H A^2 \Delta/M +SH^2 +M'H^2
\end{eqnarray}

After symmetry breaking this leads to a matrix of the following form
among the three pairs of weak doublets in the theory i.e.
$H_{u,10}, H_{u,\Delta},H_{u,\bar{\Delta}}$ and thew corresponding $H_d$'s.
In the basis where the column is given by$(10, 126, \bar{126})$ and similarly 
for the row, we have the Doublet matrix:
\begin{eqnarray}
M_D=\left(\begin{array}{ccc}
0 & <A>^2/M & 0\\
<A>^2/M & 0 & M\\
0 & M & 0\end{array}\right)
\end{eqnarray}
where the direct $H_{u,10}H_{d,10}$ mass term is fine tuned to zero. This 
kind of a three by three mass matrix leaves the low energy doublets to
have components from both the {\bf 10} and {\bf 126} and thus avoid the
mass relations. It is easy to check that the triplet mass matrix in this case
makes all of them heavy.

There is another way to achieve the similar result without resorting to
fine tuning as we did here by using {\bf 16} Higgses. Suppose there are
two {\bf 10}'s, one pair of {\bf 16} and $\bar{\bf 16}$ (denoted by
$\Psi_H,\bar{\Psi}_H$). Let us write the following superpotential:
\begin{eqnarray}
W_{bm}=\Psi_H\Psi_H H_1 +\bar{\Psi}_H\bar{\Psi}_H H_2 +AH_1H_2
+\Psi_2\Psi_2 AA'H_2
\end{eqnarray}
If we now give vev's to $<\nu^c>\neq 0$ and $\bar{<\nu^c>}\neq 0$, then
the three by three doublet matrix involving the $H_{u}$'s from $H_i$ and
$\bar{\Psi}_H$ and $H_d$'s from the $H_i$'s and $\Psi_H$ form the
three by three matrix which has the same as in the above equation.
As a result, the light MSSM doublets are admixtures of doublets from
$\bf 10$'s and $\bf 16$'s. This in conjunction with the last term in
the above superpotential gives precisely the GJ mass matrices without
effecting the form of the up quark mass matrix.

Thus it is possible to have D-T-S along with phenomenologically viable
mass matrices for fermions.

\bigskip

\subsection{Final comments on SO(10)}
\bigskip

The SO(10) model clearly has a number of attractive properties over
the SU(5) model e.g. the possibility to have automatic R-parity
conservation, small nonzero neutrino masses, interesting fermion mass
relations etc. There is another aspect of the model that makes
it attractive from the cosmological point of view. This has to do with
a simple mechanism for baryogenesis. It was suggested by Fukugita and
Yanagida\cite{fuku} that in the SO(10) type models, one could
first generate a lepton asymmetry at a scale of about $10^{11}$ GeV or 
so when the righthanded Majorana neutrinos have mass and generate
the desired lepton asymmetry via their decay. This lepton aymmetry
in the presence of sphaleron processes can be converted to baryons.
This model has been studied quantitatively in many papers and found to
provide a good explanation of the observed $n_B/n_{\gamma}$\cite{jung}.

\bigskip

\section{Other grand unifcation groups}

While the SU(5) and SO(10) are the two simplest grand unification groups,
other interesting 
 unfication models motivated for different reasons are those based
on $E_6$, $SU(6)$, $ SU(5)\times U(1)$ and $SU(5)\times SU(5)$. We discuss
 them very briefly in this final section of the lectures.

\bigskip

\subsection{$E_6$ grand unification}

These unification models were considered\cite{gursey}
in the late seventies and their popularity increased in the late eighties
after it was demonstrated that the Calabi-Yau compactification of
the superstring models lead to the gauge group $E_6$ in the visible
sector and predict the representations for the matter and Higgs multiplets
that can be used to build realistic models\cite{witten}.

To start the discussion of $E_6$ model building, let us first note
that $E_6$ contains the subgroups (i) $SO(10)\times U(1)$;
(ii) $SU(3)_L\times SU(3)_R\times SU(3)_c$ and (iii) $SU(6)\times SU(2)$.
The $[SU(3)]^3$ subgroup shows that the $E_6$ unification is also
left right symmetric.
The basic representation of the $E_6$ group is {\bf 27} dimensional and
for model building purposes it is useful to give its decomposition
interms of the first two subgroups:
\begin{eqnarray}
SO(10)\times U(1)::{\bf 27}={\bf 16}_1+{\bf 10}_{-2}+{\bf 1}_4\\ \nonumber
[SU(3)]^3::{\bf 27}= {\bf (3,1,3)+(1,\bar{3},\bar{3})+(\bar{3}, 3, 1)}
\end{eqnarray}

The fermion assignment can be given in the $[SU(3)]^3$ basis as follows:
\begin{eqnarray}
{\bf(3,1,3)}=\left(\begin{array}{c} u\\ d\\ D\end{array}\right);
{\bf (1,\bar{3},\bar{3})}=\left(\begin{array}{c} u^c\\d^c\\D^c
\end{array}\right);\\ \nonumber
{\bf (\bar{3}, 3, 1)}=\left(\begin{array}{ccc}
H^0_1 & H^+_2 & e^+\\
H^-_1 & H^0_2 & \nu^c\\
e^- & \nu & n_0  \end{array}\right)
\end{eqnarray}
We see that there are eleven extra fermion fields than the SO(10) model.
Thus the model is non minimal in the matter sector. Important to note
that all the new fermions are vector like. This is important from the low 
energy point of view since the present electroweak data\cite{alta} (i.e.
the precision measurement of radiative parameters S, T and U put severe
restrictions on extra fermions only if they are not vectorlike. Also
the vectorlike nature of the new fermions keeps the anomaly cancellation
of the standard model.

Turning now to symmetry breaking, we will consider two interesing chains-
although $E_6$ being a group of rank six, there are many possible ways to
arrive at the standard model. One chain is:
\begin{eqnarray}
E_6\rightarrow [SU(3)]^3\rightarrow G_{2213}\rightarrow G_{STD}
\end{eqnarray}
The first stage of the breaking can be achieved by a {\bf 650} dimensional
Higgs field which is the lowest dim. representation that has a singlet
under this group. In the case of string models this stage is generally
achieved by the Wilson loops involving the gauge fields along the compactified
direction. The second stage is achieved by means of the $n_0$ field in
the {\bf 27} dimensional Higgs boson. The final stage can be achieved in one
of two ways depending on whether one wants to maintain the R-parity
symmetry after symmetry breaking. If one does not care about breaking
R-parity, the $\nu^c$ field in {\bf 27}-Higgs can be used to arrive at
the standard model On the other hand if one wants to keep R-parity
conserved, the smallest dimensional Higgs field would be {\bf 351'}
is needed to arrive at the standard model.

Another interesting chain of symmetry breaking is:
\begin{eqnarray}
E_6\rightarrow SO(10)\times U(1)\rightarrow G_{2213}\rightarrow G_{STD}
\end{eqnarray}
The first stage of this chain is achieved by a {\bf 78} dim. rep. and the
rest can be achieved by the {\bf 27} Higgs as in the previous case.

The fermion masses in this model arise from {\bf 27} higgs since
${\bf 27_m 27_m 27_H}$ is $E_6$ invariant and it contains the 
MSSM doublets (the $H_i$ fields in the {\bf 27} given above.
The $[27]^3$ interaction in terms of the components can be written as
\begin{eqnarray}
[27]^3\rightarrow QQD+Q^cQ^cD^c+QQ^cH+LL^cH\\ \nonumber
+H^2n_0+DD^cn_0+QLD^c+Q^cL^cD
\end{eqnarray}
Form this we see that in addition to the usual assignments of B-L to
known fermions, if we assign $B-L$ for D as -2/3 and $D^c$ as +2/3,
then all the above terms conserve R-parity prior to symmetry breaking.
However when $<\nu^c>\neq 0$, $d^c$ a D mix leading to breakdown of R-parity.
They can for instance generate a $u^cd^cd^c$ term with strength
$\frac{<\nu^c>}{<n_0>}$ . This can lead to the $\Delta B=2$ processes
such as neutron-antineutron oscillation.

\bigskip
\subsection{$SU(5)\times SU(5)$ unification}

The $SU(5)\times SU(5)$ model that we will discuss here was motivated
by the goal of maintaining automatic R-parity conservation as well as the
simple see-saw mechanism for neutrino masses in the context of superstring
compactification. The reason was the failure of the string models at any
level to yield the {\bf 126} dim. rep. in the case of SO(10) yielding
fermionic compactifications. Although no work has been done on higher
level string compactifications with $SU(5)\times SU(5)$ as the GUT group,
the model described here involves simple enough representaions that
it may not be unrealistic to expect them to come out of a consistent
compactification scheme. In any case for pure $SU(5)$ at level II
all representations used here come out. Let us now see some details of the 
model.

The matter fields in this case belong to left-right symmetric representations
 such as ${\bf  (\bar{5},1)+(1,{5})+(10, 1)+(1, \bar{10})}$
as follows: (denoted by $F_L,F_R,T_L,T_R$).
\begin{eqnarray}
F_L=\left(\begin{array}{c} D^c_1\\ D^c_2\\D^c_3\\ e^-\\ \nu\end{array}\right);
F_R=\left(\begin{array}{c} D_1\\D_2\\D_3\\ e^+\\ \nu^c\end{array}\right)\\
\nonumber
T_L=\left(\begin{array}{ccccc}
0 & U^c_3 & -U^c_2 & u_1 & d_1\\
 & 0 & U^c_1 & u_2 & d_2\\
 &  & 0 & u_3 & d_3\\
 &   &  & 0 & E^+\\
& & & & 0\end{array}\right);\\ \nonumber
T_R=\left(\begin{array}{ccccc}
0 & U_3 & -U_2 & u^c_1 & d^c_1\\
 & 0 & U_1 & u^c_2 & d^c_2\\
 & & 0 & u^c_3 & d^c_3 \\
 & & & 0 & E^-\\
& & & & 0\end{array}\right)
\end{eqnarray}
This left-right symmetric fermion assignment was first considered in Ref.
\cite{wali}. But the R-parity conserving version of the model was considered
in Ref.\cite{rabi}. Crucial to R-parity conservation is the nature of the
Higgs multiplets in the theory. We choose the higgses belonging to
${\bf (5,\bar{5}), (15, 1) + (1, \bar{15})}$. The $ SU(5)\times SU(5)$
group is first broken down to $SU(3)_c\times SU(2)_L\times SU(2)_R\times
 U(1)_{B-L}$ by the ${\bf (5,\bar{5})}$ acquiring vev's along
$Diag(a,a,a,0,0)$. This also makes the new vectorlike particles $U,D,E$
superheavy. The left-right group is then broken down to the $G_{STD}$
by the {\bf 15}-dimensional Higgs acquiring a vev in its right handed
multiplet along the $\nu^c\nu^c$ direction. This component has $B-L=2$ and 
therefore R-parity remains an exact symmetry. The light fermion masses
and the electroweak symmetry breaking arise via the vev of a second
${\bf (5,\bar{5})}$ multiplet acquiring vev along the direction
$Diag(0,0,0,b,b)$.

A new feature of these models is that due to the presence of new fermions,
the normalization of the hypercharge and color are different from the
standard SU(5) or SO(10) unification models. In fact in this case,
$I_Y=\sqrt{\frac{3}{13}}(Y/2)$ and as result, at the GUT scale
$sin^2\theta_W=\frac{3}{16}$. The GUT scale in this case is therefore
much lower than the standard scenarios discussed prior to this.

\bigskip
\subsection{Flipped SU(5)}

This model was suggested in Ref.\cite{anto} and have been extensively
studied as a model that emerges from string compactification. It is
based on the gauge group $SU(5)\times U(1)$ and as such is not a 
strict grand unification model. Nevertheless it has several interesting
features that we mention here.

The matter fields are assigned to representations ${\bf \bar{5}_{-3}}$ (F),
${\bf {1}_{+5}}$ (S) and ${\bf 10_{+1}}$ (T).
 The deatailed particle assignments
are as follows:
\begin{eqnarray}
F=\left(\begin{array}{c} u^c_1\\u^c_2\\u^c_3\\e^-\\ \nu\end{array}\right);
T=\left(\begin{array}{ccccc}
0 & d^c_3 & -d^c_2 & u_1 & d_1\\
 & 0 & d^c_1 & u_2 & d_2 \\
 & & 0 & u_3 & d_3 \\
  &  &  & 0 & \nu^c \\
  &  &  &  &  0\end{array}\right);
S= e^+
\end{eqnarray}

The elctric charge formula for this group is given by:
\begin{eqnarray}
Q=I_{3L} -\frac{1}{\sqrt{15}}\lambda_{24} +\frac{1}{5} X
\end{eqnarray}
where $\lambda_a$ ( a= 1...24) denote the SU(5) generators and $X$ is the
$U(1)$ generator with $I_{3L}\equiv \lambda_3$. The Higgs fields are
assigned to representations $\Sigma ({\bf 10}_{+1})$ + $\bar{\Sigma}$
and $H({\bf 5}_{-2})$ + $\bar{H}$. The first stage of the symmetry breaking
in this model is accomplished by $\Sigma_{45}\neq 0$. This leaves the
standard model group as the unbroken group. Another point is that since the
$\Sigma_{45}$ has $B-L=1$, this model breaks R-parity (via the 
nonrenormalizable interactiions). $H$ and $\bar H$ contain the MSSM
doublets. An interesting point about the model is the natural way
in which doublet-triplet splitting occurs. To see this note the most general 
superpotential for the model involving the Higgs fields:
\begin{eqnarray}
W_5=\epsilon_{abcde} \Sigma^{ab}\Sigma^{cd} H^e + \epsilon^{abcde}
\bar{\Sigma}_{ab}\bar{\Sigma}_{cd}\bar{H}_e
\end{eqnarray}
On setting $\Sigma_{45}= M_U$, the first term gives $\epsilon_{ijk}\Sigma^{ij}
H^k$ which therefore pairs up the triplet in $H$ with the triplet in
${\bf 10}$ to make it superheavy and since there is no color singlet
weak doublet in {\bf 10}, the doublet remains light . This provides a neat
realization of the missing partner mechanism for D-T-S.

The fermion masses in this model are generated by the following superpotential:
\begin{eqnarray}
W_F= h_d TTH + h_u T\bar{F}\bar{H} + h_e \bar{F}H S
\end{eqnarray}
It is clear that this model has no $b-\tau$ mass unification; thus we 
lose one very successful prediction of the SUSY GUTs. There is also no simple
 see-saw mechanism. And furthermore the model does not conserve R-parity 
automatically as already noted. For instance there are higher dim.
terms of the form $TT\Sigma\bar{F}/M_{Pl}$, $\bar{F}\bar{F}\Sigma S/M_{Pl}$
that after symmetry breaking lead to R-parity breaking terms like $QLd^c$
and $LLe^c$. Thus they erase the baryon asymmetry in the model.

\bigskip
\subsection{SU(6) GUT and naturally light MSSM doublets:}

In this section, we discuss an SU(6) GUT model which has the novel 
feature that under certain assumptions the MSSM Higgs doublets arise
as pseudo-Goldstone multiplets in the process of symmetry breaking
without any need for fine tuning. This idea was suggested by Berezhiani and 
Dvali\cite{bere} and has been pursued in several subsequent papers\cite{lisa}.

We will only discuss the Higgs sector of the model since our primary
goal is to illustrate the new mechanism to understand the D-T-S.
Consider the Higgs fields belonging to the {\bf 35} (denoted by $\Sigma$),
and to {\bf 6} and ${\bf {\bar{6}}}$ (denoted by $H,\bar{H}$ respectively).
Then demand that the superpotential of the model has the following
structure:
\begin{eqnarray}
W= W_{\Sigma} + W(H,\bar{H})
\end{eqnarray}
i.e. set terms such as $H\Sigma \bar{H}$ to zero. This is a rather adhoc
assumption but it has very interesting consequences. Let the fields have
the following pattern of vev's.
\begin{eqnarray}
<H>=<\bar{H}>=\left(\begin{array}{c} 1 \\0\\0\\0\\0\\ 0\end{array}\right);
<\Sigma>= Diag (1,1,1,1,-2,-2)
\end{eqnarray}
Note that $\Sigma$ breaks the SU(6) group down to $SU(4)\times SU(2)\times 
U(1)$ whereas $H$ field breaks the group down to $SU(5)\times U(1)$.
Note that the Goldstone bosons for the breaking to $SU(5)\times U(1)$ are
in ${\bf 5+\bar{5}+1}$ i.e. under the standard model group they tranform
as : ($SU(3)\times SU(2)\times U(1)$)
\begin{eqnarray}
GB's= {\bf (3,1)+(1,2)+(\bar{3},1)+(1,2)+(1,1)}
\end{eqnarray}
whereas the Goldstone bosons generated by the breaking of $ SU(6)\rightarrow
SU(4)\times SU(2)\times U(1)$ by the $\Sigma$ are:
\begin{eqnarray}
{\bf (\bar{3},2)+(3,2)+(1,2)+(1,2)}
\end{eqnarray}
Since both the vev's break $SU(6)\rightarrow SU(3)\times SU(2)\times U(1)$,
the massless states that are eaten up in the process of Higgs mechanism
are
\begin{eqnarray}
{\bf (3,1)+(\bar{3},1) +(1,2)+(1,2)+(3,2)+(\bar{3},2)+(1,1)}
\end{eqnarray}
We then see that the only Goldstones that are not eaten up are the two weak
doublets {\bf (1,2)+ (1,2)}. These can be identified with the MSSM doublets.
This model can be made realistic by adding matter fermiona to two 
${\bf \bar{6}}$'s and a {\bf 15} per generation to make the model anomaly
free. We do not discuss this here.

\section{Epilogue}

This set of lectures is meant to be a pedagogical overview of the vast (and 
still expanding) field of supersymmetric grand unification. The 
body of literature
in this field is large and only a very selective sample has been given.
They should be consulted for additional references.

While this is a very interesting field, it is by no means clear that
a simple GUT group is the only way to achieve unification of particles
and forces in the universe. It could for instance be that at the string
scale, in superstring theories, the standard model or an extended version of it
with extra U(1)'s emerges directly. This possibility has certain advantages
and distinct signatures. For instance, one need not worry about questions
such as doublet-triplet splitting in such models and there would be no
need for proton to decay. There is however a puzzle with this scenario-
i.e. the MSSM spectrum leads to unification around $10^{16}$ GeV whereas
the string scale is around $10^{17.6}$ GeV or so. How does one understand
this gap. It could be that there are intermediate scales or new particles
that change the running of couplings that close this gap. A more intriguing
suggestion recently put forward by Horava and Witten\cite{horava} is that
the existence of an 11th dimension with an appropriate compactification
along it might change the running of the gravitational coupling from
naive $G_NE^2$ to $G_NE^3(R_{11})$ (where $R_{11}$ is the compactification
radius of the 11th dimension) in such a way that it might bring all
couplings to unify at the same $10^{16}$ GeV scale.

In the early days of grand unification, it used to be thought that in
addition to the attractive property of unification of couplings, the
GUT models are needed for an understanding of electric charge quantization and
the origin of matter of in the universe. It is now known that one can
understand the electric charge quantization using only cancellation of
gauge anomalies; moreover, while GUT models lead to quantization of electric
charge, to obtain observed values of these charges, an extra assumption 
regarding the Higgs representations is needed. Thus understanding the
values of the electric charges of elementary fermions needs more than
a simple GUT group. 

On the cosmological side,
the advent of the idea of weak scale baryogenesis has largely overshadowed
the significance of grand unification in understanding the baryon 
asymmetry of nature. Thus ultimately, the unification of coupling constants
may be the only (though very attractive) motivation for grand unification.
This paragraph is meant to convey the sentiment that grand unification should
not be considered a penacea for all the woes of the standard model but
as an interesting approach to a more elegant extension. 

On more phenomenological level, tests of the grand unification idea are
always quite model dependent;  if any of them show up, we will
know that the idea may be operative whereas if no experimental signal
appears, it will not necessarily rule out the idea or make it any less
plausible. An analogy may be made with the corresponding situation in 
supersymmetry. Most people believe that if the standard model Higgs 
boson with a
mass less than 150 GeV does not show up at the LHC or some other high
energy machine, interest in supersymmetry as an idea relevant for physics
will lessen considerably. There is no such stringent test for SUSY GUTs.
On the other hand, observation significant flavor violation as in
$\mu\rightarrow e+\gamma$\cite{barbieri} or $p\rightarrow K^+\nu_{mu}$
or $N-\bar{N}$ oscillation will signal some form of grand unification.
There will then be an urgency to focus on particular GUT models and to
solve the various problems associated with them.

On the theoretical side, understanding the fermion mass and mixing
hierarchies may or may not suggest SUSY GUT. While SUSY GUTs provide
one class of models for this discussion, the fermion mass problem
could also be addressed  within the framework of radiative corrections
as in the examples discussed in Ref.\cite{bala}. Then there are the
recent indications of neutrino masses from various experiemnts such
as the solar and atmospheric neutrino expaeriments. If they are
confirmed, they will certainly be strong indications of a local
B-L symmetry as well as left-right symmetric grand unification and
a scale of these new symmetries most likely in the $10^{11}$ Gev range.

Finally, the interplay between the hidden sector and the SUSY GUT of
flavor is an interesting venue for research. Could complex structures
for the hidden and the messenger sectors be maintained without sacrificing
unification of couplings. What is the role of superstring theories in
dictating the hidden sector ? Is it the hidden sector gluino condensate
that plays the role of the Polonyi singlet or is it different as in the 
anomalous U(1) models ? SUSY GUTs with anomalous U(1) remains essentially
unexplored and more work is needed to unravel its full ramifications.

The field clearly has immense possibilities and hopefully this review
will provide a summary of the relevant basic ideas that a beginner could
use to make effective contributions which are so badly needed in so
many areas.

I would like to thank Jon Bagger and K. T. Mahanthappa for kind
invitation to present these lectures at TASI97 and warm hospitality
in a pleasant Boulder summer, the 
students for their responses and comments and many colleagues for sharing 
their insight and enthusiasm into various aspects of supersymmetric
grand unification. In particular, I wish to thank K. S. Babu,
Z. Berezhiani, B. Brahmachari, Z. Chacko, B. Dutta, R. Kuchimanchi,
S. Nandi, M. K. Parida, J. C. Pati, A. Rasin, A. Riotto, G. Senjanovi\'c
for discussions over the years on many of the topics discussed here.
This work has been supported by the National Science
Foundation grant no. PHY-9421386.


\newpage                                                           
                                                                          
\section*{References}                                                 
\begin{thebibliography}{99}                                          
                                                     
\bibitem{ss|bagger} J Bagger, J Wess: {\it Supersymmetry and
Supergravity}, Princeton University Press (1983).

\bibitem{ss|mohap} R N Mohapatra: {\it Unification and Supersymmetry},
Springer-Verlag, Second edition (1991).

\bibitem{ss|HaKa84} H Haber, G Kane:  Phys. Rep. {\bf 117}, 76 (1984).

\bibitem{ss|Nil84} H P Nilles:  Phys. Rep. {\bf  110}, 1 (1984).

\bibitem{ss|GRS79} M Grisaru, M Rocek, W Siegel: Nucl. Phys. {\bf B159},
429 (1979).

\bibitem{ss|susy96} {\sl Supersymmetry-96: Theoretical
Perspectives and Experimental Outlook}, ed. R N Mohapatra and A
Rasin, North Holland (1997).

\bibitem{howie} H. Haber, TASI lectures (1994); for a recent review on
                                                                        
the open questions in supersymmetric particle physics, see K. Dienes and
C. Kolda, hep-ph/9712322.

\bibitem{dawson} S. Dawson, these proceedings.

\bibitem{ss|polonyi} R. Arnowitt, A. Chamseddine and P. Nath, {\it
Applied N=1
Supergravity}, World Scientific (1984); R. Barbieri, S. Ferrara and
C. Savoy, Phys. Lett. {\bf 119B}, 343 (1982); J. Polonyi, Budapest
preprint, KFK-1977-93 (1977).


\bibitem{ss|dine} M Dine, A Nelson: {\bf D 48}, 1277 (1993);
 M. Dine, A.E. Nelson, Y. Nir, and Y. Shirman,
 hep-ph/9507378;  Phys. Rev. {\bf D53} (1996) 2658; A.E. Nelson,
hep-ph/9511218;
M. Dine and A.E. Nelson,  Phys. Rev. {\bf D48}, 1277 (1993);
M. Dine, A.E. Nelson, and Y. Shirman,  Phys. Rev. {\bf D51}, 1362 (1995).


\bibitem{ss|binetruy} P Binetruy, E Dudas, Phys. Lett. {\bf B389}, 503 
(1996); G Dvali, A Pomarol, Phys. Rev. lett. {\bf 77}, 3728 (1996);
R N Mohapatra, A Riotto, Phys. Rev. {\bf D55}, 4262 (1997).

\bibitem{ss|aul82} C. S. Aulakh and R. N. Mohapatra, Phys. Lett. {\bf 119B},
36 (1982); L. Hall and M. Suzuki, Nucl. Phys. {\bf B231}, 419 (1984);
V Barger, G F Giudice, M Y Han; Phys. Rev. {\bf 40}, 2987 (1989);
For a recent review, see G Bhattacharyya, Proceedings of SUSY'96,
Nucl. Phys. (Proc. Suppl.), {\bf 52A}, 83 (1996).

\bibitem{ss|masiero} F. Gabbiani, E. Gabrielli, A. Masiero and L.
Silvestrini, hep-ph/9604387.

\bibitem{garisto} For a review see R. Garisto, Nucl. Phys. {\bf B419}, 279
(1994).


\bibitem{ss|gira} E. Cremmer, S. Ferrara, L. Girardello and A. van Proeyen,
Nucl. Phys. {\bf B212}, 413 (1983); J. Bagger, Nucl. Phys. {\bf B211}, 302
(1983).

\bibitem{ss|alva} K. Inoue et al. Prog. Theor. Phys. {\bf 68}, 927 (1982);
L. Ibanez and G. Ross, Phys. Lett. {\bf B110}, 227 (1982); L.
Alvarez-gaume, M. Polchinski and M. Wise, Nucl. Phys. {\bf B250}, 495
(1983).

\bibitem{nano} For a review see, E. Cremmer, S. Ferrara, C. Kounas and
D. V. Nanopoulos, Phys. Lett. {\bf 133B}, 61 (1983).

\bibitem{witten1} E. Witten, Phys. Lett. {\bf 155B}, 151 (1985).
                                                     
\bibitem{dutta}  R. N. Mohapatra and S. Nandi, Phys. Rev. Lett. {\bf 79},
181 (1997); Z. Chacko, B. Dutta, R. N. Mohapatra and S. Nandi, Phys. Rev.
{\bf D 56}, 5466 (1997).

\bibitem{GMSB} S. Ambrosanio, G. Kane, G. Kribs, S. Martin and S. Mrenna,
Phys. Rev. Lett. {\bf 76} (1996) 3498; S. Dimopoulos, M. Dine, S. Raby
and S. Thomas, {\it ibid} {\bf 76} (1996) 3494; K. S. Babu, C.
Kolda and F. Wilczek,
hep-ph/9605408; S. P. Martin, hep-ph/9608224  ; G. Dvali,
G. F. Giudice and A.
Pomarol, hep-ph/9603238; A. Riotto, N. Tornquist and R. N.
Mohapatra,
Phys. Lett. {\bf B 388}, 599 (1996); S. Dimopoulos, S.
Thomas and J. Wells,
hep-ph/9609434; J. Bagger, D. Pierce, K. Matchev and R.-J
Zhang,
hep-ph/9609443; E. Poppitz and S. Trivedi, hep-ph/9609529;
I. Dasgupta,
B. Dobrescu and L. Randall, hep-ph/9607487; H. Baer et al,
hep-ph/9610358;
A. de Gouvea, T. Moroi and H. Murayama, hep-ph/9701244;
D. Dicus, B. Dutta and S. Nandi, hep-ph/9701341; N.
Arkani-hamed, H.
Murayama and J. March-Russell, hep-ph/9701286; G. Bhattacharyya and
A. Romanino, hep-ph/9611243;  S. Raby, Phys. Rev. {\bf D56}, 2852 
(1997);  M. Luty, hep-ph/9706554.

\bibitem{dine1} M. Dine, N. Seiberg and E. Witten, Nucl. Phys. {\bf B289},
585 (1987); J. Attick, L. Dixon and A. Sen, {\it ibid} {\bf B292}, 109
(1987).

\bibitem{anom} G. Dvali and A. Pomarol, hep-ph9708304; Z. Berezhiani and
Z. Tavartkiladze, Phys. Lett. {\bf B396}, 150 (1997) and {\bf B409}, 220 
(1997);  Ren-Jie Zhang, hep-ph/9702333; P.
Binetruy, N. Irges, S. Lavignac and P. Ramond, hep-ph/9612442; N. Arkani-
Hamed and H. Murayama, hep-ph/9703259; A. Nelson and D. Wright,
hep-ph/9702359; A. Faraggi and J. C. Pati, hep-ph/9712516.

\bibitem{ss|Gid88} S Giddings, A Strominger, Nucl. Phys.
{\bf B 307}, 854 (1988).
                                                     
\bibitem{ss|Moh86} R N Mohapatra: Phys. Rev. {\bf 34}, 3457 (1986);
A. Font, L. Ibanez and F. Quevedo, Phys. Lett. {\bf B228}, 79 (1989);
S. P. Martin, Phys. Rev. {\bf D46}, 2769 (1992).

\bibitem{ramond} M. Gell-Mann, P. Ramond and R.
Slansky, {\it Supergravity}, ed. D. Freedman and P.
van Niuenhuizen (North Holland, 1979); T. Yanagida,
KEK Lectures, (1979); R. N. Mohapatra and G.
Senjanovi\'c, Phys. Rev. Lett. {\bf 44}, 912 (1980).

\bibitem{ss|kuchi} R Kuchimanchi, R N Mohapatra: Phys. Rev. {\bf 48}, 4352
(1993).

\bibitem{ss|kuchi2} R. Kuchimanchi and R. N. Mohapatra, Phys. Rev. Lett.
{\bf 75}, 3989 (1995); C. Aulakh, K. Benakli and G. Senjanovi\'c,
hep-ph/9703434; Phys. Rev. Lett. {\bf 79}, 2188 (1997).

\bibitem{goran} R. N. Mohapatra and A. Rasin, Phys. Rev. Lett. {\bf 76},
3490 (1996); C. S. Aulakh, A. Melfo and G. Senjanovi\'c, hep-ph/9707256;
C. S. Aulakh, A. Melfo, A. Rasin and G. Senjanovi\'c, hep-ph/9712551.

\bibitem{chacko} Z. Chacko and R. N. Mohapatra, hep-ph/9712359.

\bibitem{rasin2} R. N. Mohapatra and A. Rasin, Phys. Rev. {\bf D54}, 5835
(1996).

\bibitem{kmr} R. N. Mohapatra and A. Rasin, Phys. Rev. Lett. {\bf 76}, 
3490 (1996); R. Kuchimanchi, Phys. Rev. Lett. {\bf 76}, 3486 (1996);
R. N. Mohapatra, A. Rasin and G. Senjanovi\'c, Phys. Rev. Lett. {\bf 79},
4744 (1997).

\bibitem{posp} M. Pospelov, Phys. Lett. {\bf B391}, 324 (1997).

\bibitem{ss|FFK91} M Francis, M Frank, C S Kalman, Phys. Rev.
{\bf D 43}, 2369 (1991);
K Huitu, J Malaampi, M Raidal, Nucl. Phys.
{\bf B 420}, 449 (1994).

\bibitem{pati} J. C. Pati and A. Salam, Phys. Rev. {\bf D10}, 275 (1974).

\bibitem{georgi} H. Georgi and S. L. Glashow, Phys. Rev. Lett. {\bf 32},
438 (1974).

\bibitem{quinn} H. Georgi, H. Quinn and S. Weinberg, Phys. Rev. Lett.
{\bf 33}, 451 (1974).

\bibitem{many} W. Marciano and G. Senjanovi\'c, Phys. Rev. {\bf D25},
3092 (1982);
U. Amaldi, W. de Boer and H. Furstenau, Phys. Lett. {\bf B260},447 (1991);
P. Langacker and M. Luo, Phys. Rev. {\bf D44}, 817 (1991); J. Ellis,
S. kelly and D. Nanopoulos, Phys. Lett. {\bf B260}, 131 (1991).

\bibitem{parida} D. Chang, R. N. Mohapatra, M. K.
Parida, J. Gipson and R. E. Marshak, Phys. Rev. {\bf
D31}, 1718 (1985); N. G. Deshpanda, E. Keith and P.  
Pal, Phys. Rev. {\bf D46}, 2261 (1992); R. N.
Mohapatra and M. K. Parida, Phys. Rev. {\bf D 47},
264 (1993); F. Acampora, G. Amelino-Camelia, F.
Buccella, O. Pisanti, L. Rosa and T. Tuzi, DSF
preprint 93/52.

\bibitem{brahma} B. Brahmachari and R. N. Mohapatra, Int. Journ. Mod. Phys.
{\bf A 11}, 1699 (1996).

\bibitem{schmelling} M. Schmelling, hep-ex/9701002.

\bibitem{mar} Many papers have addressed this issue; see for instance,
M. Bastero-gil and J. Perez-Mercader, Nucl. Phys. {\bf B450}, 21 (1995);
P. Langacker and N. Polonsky, Phys. Rev. {\bf D 52}, 3081 (1995).

\bibitem{bagger} J. Bagger, K. Matchev, D. Pierce and R. J. Zhang,
hep-ph/9611229 and hep-ph/9606211.

\bibitem{chan} M. Olechowski and S. Pokorski, Phys. Lett. {\bf B214}, 393
(1988); B. Ananthanarayan, G. Lazaridis and Q. Shafi, Phys. Rev. {\bf D 44},
1613 (1991); R. Rattazi, U. Sarid and L. hall, hep-ph/9405313.

\bibitem{barger} V. Barger, M. Berger, T. Han and M. Zralek, Phys.
Rev. Lett. {\bf 68}, 3894 (1992).

\bibitem{hill} C. Hill, Phys. Rev. {\bf D24}, 691 (1981);
C. Penddleton and G. Ross, Phys. Lett. {\bf B98}, 291 (1981).

\bibitem{dimo} S. Dimopoulos and H. Georgi, Nucl. Phys. {\bf B193}, 150
(1981); N. Sakai, Z. Phys. {\bf C11}, 153 (1981).

\bibitem{nath2} R. Arnowitt, A. Chamsheddine and P. Nath, Phys. Lett.
{\bf B156}, 215 (1985).

\bibitem{nath3} For a review see, R. Arnowitt and P. Nath, {\it Lectures
at the VII J. Swieca Summer school, 1993}, CTP-TAMU-52/93.

\bibitem{zura} R. Barbieri, Z. Berezhiani and A. Strumia, hep-ph/9704275.

\bibitem{kuzmin} V. Kuzmin, V. Rubakov and M. Shaposnikov, Phys. Lett.
{\bf B155}, 36 (1985); for a review and other references, see A. Cohen,
D. Kaplan and A. Nelson, Ann. Rev. Nucl. Sc. {\bf 43}, 27 (1993).

\bibitem{sakita} R. N. Mohapatra and B. Sakita, Phys. Rev. {\bf D21},
1062 (1980).

\bibitem{zee} For an alternative formulation see, F. Wilczek and A. Zee,
Phys. Rev. {\bf D25}, 553 (1982).


\bibitem{parida} D. Chang, R. N. Mohapatra and M. K. parida, Phys. Rev. Lett.
{\bf 52}, 1072 (1974).

\bibitem{shafi} T. Kibble, G. Lazaridis and Q. Shafi, Phys. Rev. {\bf D26},
435 (1982).

\bibitem{jarl} H. Georgi and C. Jarlskog, Phys. Lett. {\bf B86}, 297 (1979).

\bibitem{he} J. Harvey, P. Ramond and D. Reiss, Phys. Lett. {\bf 92B},
309 (1980);
 X. G. He and S. Meljanac, Phys. Rev. {\bf D41}, 1620 (1990);
S. Dimopoulos L. Hall and S. Raby, Phys. Rev. Lett. {\bf 68}, 1984
(1992); K. S. Babu and R. N. Mohapatra, Phys. Rev. Lett. {\bf 74},
2418 (1995).

\bibitem{pal} R. N. Mohapatra and P. B. Pal, {\it Massive neutrinos in
 Physics and Astrophysics}, (World Scientific, 1989).

\bibitem{dienes} K. Dienes, Phys. Rep. {\bf 287}, 447 (1997);
K. Dienes and J. March-Russel, hep-th/9604112; see however, J. Erler,
Nucl. Phys. {\bf B475}, 597 (1996) where it is argued one may get {\bf 126}'s
in more general string compactifications.

\bibitem{lykken} S. Choudhuri, S. Chung and J. Lykken, hep-ph/9511456;
G. Cleaver,hep-th/9708023; S. Choudhuri, S. Chung, G. Hockney and J.
Lykken, Nucl. Phys. {\bf 456}, 89 (1995); G. Aldazabal, A. Font, L.
Ibanez and A. M. Uranga, Nucl. Phys. {\bf B452}, 3 (1995).

\bibitem{hall} K. S.Babu and S. Barr, Phys. Rev. {\bf D50}, 3529 (1994);
K. S. Babu and R. N. Mohapatra, Phys. Rev. Lett. {\bf 74}, 2418 (1995);
L. Hall and S. Raby, Phys. Rev. {\bf D51}, 6524 (1995); D. G. Lee and
R. N. Mohapatra, Phys. Rev. {\bf D51}, 1353 (1995); S. Barr and S. Raby,
hep-ph/9705366; C. Albright and S. Barr, hep-ph/9712488; Z. Berezhiani,
Phys. Lett. {\bf B355}, 178 (1995).

\bibitem{valle} R. N. Mohapatra, Phys. Rev. Lett. {\bf 56}, 561 (1986);
R. N. Mohapatra and J. W. F. Valle, Phys. Rev. {\bf D34}, 1642 (1986).

\bibitem{dimo} S. Dimopoulos and F. Wilczek, I.T.P preprint (unpublished);
K. S. Babu and S. Barr, Phys. Rev. {\bf D48}, 5354 (1993).

\bibitem{lee} D. G. Lee and R. N. Mohapatra, Phys. Rev. {\bf D51}, 1353
(1995).

\bibitem{babu2} K. S. Babu and R. N. Mohapatra, Phys. Rev. Lett.
{\bf 74}, 2418 (1995).

\bibitem{fuku} M. Fukugita and T. Yanagida, Phys. lett. {\bf B174}, 45 (1986).

\bibitem{jung} T. Gherghetta and G. Jungman, Phys. Rev. {\bf D 48}, 1543
(1993); W. Buchmuller and M. Plumacher, hep-ph/9711208.

\bibitem{gursey} F. Gursey and P. Sikivie, {\bf 36}, 775 (1976); P. Ramond,
Nucl. Phys.{\bf B 110}, 224 (1976).

\bibitem{witten} For a review of these developments, see M. Green,
J. Schwarz and E. Witten, {\it Superstring Theories}, (Cambridge 
University Press, 1989).

\bibitem{alta} G. Altarelli, hep-ph/9710434.

\bibitem{wali} A. Davidson and K. C. Wali, Phys. Rev. Lett. {\bf 58}, 2623
(1987); Peter Cho, Phys. Rev. {\bf D48}, 5331 (1993).

\bibitem{rabi} R. N. Mohapatra, Phys. Lett. {\bf B379}, 115 (1996).

\bibitem{anto} I. Antoniadis, J. Ellis, J. Hagelin and D. Nanopoulos,
Phys. Lett. {\bf B 231}, 65 (1989); J. Lopez, D. V. Nanopoulos and
K. Yuan, Nucl. Phys. {\bf B399}, 654 (1993).

\bibitem{bere} Z. Berezhiani and G. Dvali, Sov. Phys. Leb. Inst. Rep.
{\bf 5}, 55 (1989).

\bibitem{lisa} Z. Berezhiani, L. Randall and C. Csaki, Nucl. Phys.
{\bf B444}, 61 (1995); Z. Berezhiani, Phys. Lett. {\bf B355}, 481 (1995);
R. Barbieri, Nucl. Phys. {\bf B432}, 49 (1994); G. Dvali and S. Pokorski,
hep-ph/9610341.

\bibitem{horava} P. Horava and E. Witten, Nucl. Phys. {\bf B475}, 94
(1996).

\bibitem{barbieri} L. Hall, V. Kostelecky and S. Raby, Nucl. Phys.
{\bf B267}, 415 (1986); R. Barbieri and L. Hall, Phys. Lett. {\bf B338},
212 (1994); N. G. Deshpande, B. Dutta and E. Keith, OITS-595; T. V.
Duong, B. Dutta and E. Keith, hep-ph/9510441.

\bibitem{bala} B. Balakrishna, A. Kagan and R. N. Mohapatra, Phys. Lett.
{\bf 205B}, 345 (1988); For a review, see K. S. Babu and E. Ma, Mod.
Phys. Lett. {\bf A4}, 1975 (1989). 

\end{thebibliography}
\end{document}